\documentclass[english,apj,numberedappendix,preprint]{emulateapj}
\usepackage[T1]{fontenc}
\usepackage[latin9]{inputenc}
\usepackage[active]{srcltx}
\usepackage{babel}
\usepackage{amssymb}
\usepackage{graphicx}
\usepackage[unicode=true,pdfusetitle,
 bookmarks=true,bookmarksnumbered=false,bookmarksopen=false,
 breaklinks=false,pdfborder={0 0 1},backref=section,colorlinks=false]
 {hyperref}
\hypersetup{
 pdfcreator={Emacs Org-mode version 7.8.09}}
\usepackage{breakurl}

\makeatletter

\providecommand{\tabularnewline}{\\}

\usepackage{amsmath}
\usepackage{times}
\usepackage{url}

\makeatother

\begin{document}
\global\long\def\Mo{M_{\odot}}
\global\long\def\Ro{R_{\odot}}
\global\long\def\Lo{L_{\odot}}
\global\long\def\To{T_{\mathrm{eff,}\odot}}
\global\long\def\MKo{M_{K,\odot}}
\global\long\def\Rms{R_{\mathrm{MS}}}

\global\long\def\Mbh{M_{\bullet}}
\global\long\def\vbh{v_{\bullet}}
\global\long\def\Ms{M_{\star}}
\global\long\def\Rs{R_{\star}}
\global\long\def\Ts{T_{\star}}
\global\long\def\vs{v_{\star}}
\global\long\def\eM{\left\langle M_{\star}\right\rangle }
\global\long\def\eMM{\left\langle M_{\star}^{2}\right\rangle }
\global\long\def\s{\sigma_{\star}}
\global\long\def\n{n_{\star}}

\global\long\def\ab{a_{12}}
\global\long\def\Pb{P_{12}}
\global\long\def\Tb{T_{12}}
\global\long\def\Mb{M_{12}}
\global\long\def\Eb{E_{12}}
\global\long\def\vb{v_{12}}
\global\long\def\sb{\sigma_{12}}
\global\long\def\Lb{\Lambda_{12}}
\global\long\def\tev{t_{\mathrm{evap}}}
\global\long\def\tr{t_{\mathrm{rlx}}}
\global\long\def\mtr{\min t_{\mathrm{rlx}}}
\global\long\def\mnMM{\max\left(\n\eMM\right)}

\title{Constraining the dark cusp in the Galactic Center by long-period
binaries}

\author{Tal Alexander%
\footnote{Email: tal.alexander@weizmann.ac.il %
}}

\affil{Dept. of Particle Physics \& Astrophysics, Weizmann Institute of
Science, P.O. Box 26, Rehovot 76100, Israel }

\and

\author{Oliver Pfuhl}

\affil{Max-Planck-Institut f\"ur extraterrestriche Physik, Postfach 1312,
Garching D-85741, Germany.}
\begin{abstract}
Massive black holes (MBHs) in galactic nuclei are believed to be surrounded
by a high density stellar cluster, whose mass is mostly in hard-to-detect
faint stars and compact remnants. Such dark cusps dominate the dynamics
near the MBH: a dark cusp in the Galactic center (GC) of the Milky
Way would strongly affect orbital tests of General Relativity there;
on cosmic scales, dark cusps set the rates of gravitational wave emission
events from compact remnants that spiral into MBHs, and they modify
the rates of tidal disruption events, to list only some implications.
A recently discovered long-period massive young binary  ($\Pb\lesssim1\,\mathrm{yr}$,
$\Mb\sim{\cal O}(100\, M_{\odot})$, $\Tb\sim6\times10^{6}\,\mathrm{yr}$),
only $\sim0.1\,\mathrm{pc}$ from the Galactic MBH \citep{pfu+13},
sets a lower bound on the 2-body relaxation timescale there, $\min t_{\mathrm{rlx}}\propto(\Pb/\Mb)^{2/3}\Tb\sim10^{7}\,\mathrm{yr}$,
and correspondingly, an upper bound on the stellar number density,
$\max\n\sim\mathrm{few\times}10^{8}/\eMM\,\mathrm{pc^{-3}}$, based
on the binary's survival against evaporation by the dark cusp. However,
a conservative dynamical estimate, the drain limit, implies $t_{\mathrm{rlx}}>{\cal O}(\mathrm{10^{8}}\,\mathrm{yr})$.
Such massive binaries are thus too short-lived and tightly bound to
constrain a dense relaxed dark cusp. We explore here in detail the
use of longer-period, less massive and longer-lived binaries ($\Pb\sim\mbox{few}\,\mathrm{yr}$,
$\Mb\sim2-4\, M_{\odot}$, $\Tb\sim10^{8}-10^{10}\,\mathrm{yr}$),
presently just below the detection threshold, for probing the dark
cusp\texttt{\textbf{,}} and develop the framework for translating
their future detections among the giants in the GC into dynamical
constraints.
\end{abstract}


\section{Introduction}

The dynamical state of the stellar cluster around the massive black
hole (MBH) in the Galactic Center (GC), and the mechanisms that determine
it, are of interest because such systems are ubiquitous in the universe,
and because the GC is by far the most observationally accessible of
them. The key question is whether the GC is dynamically relaxed \citep{ale11}.
An unrelaxed system reflects its particular formation history. In
contrast, the properties of a relaxed system can be understood and
modeled from first principles, independently of initial conditions,
and extrapolated to other relaxed galaxies. The GC, like other galactic
nuclei with low-mass MBHs, is expected to be dynamically relaxed if
it evolved passively in isolation \citep{bar+13}, and consequently
to have a centrally concentrated power-law stellar cusp, with a density
profile $\propto r^{\alpha}$, where $\alpha\sim3/2-5/2$ \citep{bah+77,ale+09,kes+09}
(Figure \ref{f:IMFs}). However, recent photometric and spectroscopic
studies of the stellar distribution in the GC \citep{buc+09,do+09,bar+10}
indicate that the radial density profile of the spectroscopically
identified low-mass red giants on the $\sim0.5$ pc scale, thought
to trace the long-lived population there, rises inward much less steeply
than expected in steady state, or even decreases toward the center. 

The core in the spatial distribution of the old red giants could still
be consistent with a relaxed GC, if some mechanism selectively destroys
giants, or rejuvenates them to appear as hot stars \citep[see  review by][]{ale05}.
However, all such processes proposed to date (e.g. envelope stripping
by star-giant collisions, \citealt{ale99,bai+99}; tidal heating,
\citealt{ale+03a}) are ineffective outside the central $\sim0.1$
pc. The simplest interpretation for the missing old cusp is that,
contrary to expectations, the GC is not dynamically relaxed by two-body
interactions. One possibility is that the cusp is continuously and
rapidly drained into the MBH by faster competing dynamical processes
\citep{mad+11}, in particular resonant relaxation \citep{rau+96}.
However, a recent exact derivation of the 2-body relaxation rate \citep{ale11a,bar+13}
suggest that it is fast enough to maintain a high density steady state
cusp on the relevant spatial scales, even in the presence of resonant
relaxation. Another possibility, which may well explain the observed
cores in the light distributions of galaxies with more massive MBHs
\citep{mil+02}, is that a cosmologically recent major merger event
with another MBH carved out a central cavity in the stellar distribution
\citep{mer10}. This would then mean that the GC has not evolved in
isolation, and is not in steady state but still slowly evolving back
to it. 

There are however strong reasons to suspect that the GC is evolving
at a rate faster than that of slow 2-body relaxation. The latest star
formation episode in the GC formed $O(100)$ very massive stars \citep{pau+06},
which will likely leave behind stellar mass black holes (BHs). There
are indications that a similar star formation episode occurred $O(10^{8}\,\mathrm{yr})$
ago \citep{kra+95}, as well as further indications of continuous
star formation in the GC over a substantial fraction of the Hubble
time on various spatial scales \citep{ale+99a,fig+04,bau+10,blu+03,pfu+11}.
The estimated number of BHs produced in the inner parsec in the course
of such a star formation history, $O(10^{4})$, is in itself large
enough to lower the local relaxation time below the Hubble time \citep{ale11}.
In addition, the massive gas structures presently observed just outside
the MBH's radius of influence at $\sim2\,\mathrm{pc}$ (gas clumps
and giant molecular clouds), act as massive perturbers \citep{per+07}
and can likewise lower the relaxation time below the Hubble time even
well inside the radius of influence \citep{ale11}. Even if the GC
has undergone a cusp-scouring merger event, the re-formation rate
of the mass-segregated cusp of stellar BHs is expected to be faster
than that of the luminous, lower mass stars \citep{mer10}. In fact,
numerical experiments indicate that the presence of a mass spectrum
(not included in the analysis of \citealt{mer10}) accelerates the
dynamics to the extent that the cusp can reform in substantially less
than a Hubble time \citep{pre+10}, for a wide range of initial conditions.

The existence or absence of a dark cusp has important implications
for processes that involve strong interactions with the MBH, such
as the emission of gravitational waves from compact objects that inspiral
into it \citep[see review by][]{ama+07}, the disruption of stars
by the tidal field of the MBH and the resulting tidal flares \citep[see review by ][]{ale12},
and for the modeling of dynamics very close to MBHs \citep[e.g.][]{hop+06a}.
A dark cusp around the Galactic MBH has crucial implications for the
detection of relativistic effects in orbits of stars close to the
Galactic MBH (the so called ``S-cluster'') \citep{zuc+05,mer+10},
in particular by future high precision interferometric experiments
such as GRAVITY \citep{eis+08}. 

The discrepancy between the simplest interpretation of the observations
and the theoretical expectations, and the absence of other lines of
evidence for a major merger in the Milky Way's past, suggests that
the possibility that the observed red giants \emph{do not} trace the
total mass distribution, and that there \emph{does} \emph{exist} a
dark cusp around the Galactic MBH, should be considered and tested
empirically. The dynamical upper limits on the dark distributed mass
within the S-cluster in the inner arcsec are still at least two orders
of magnitude higher than expected from theoretical constraints \citep{gil+09}.
Direct detection of the dark cusp in the GC, for example by gravitational
lensing \citep{ale+01c,cha+01a} or by X-ray emission from accretion
\citep{pes+03}, is very difficult. The most promising approach to
detect the faint stars and compact remnants appears to be through
their dynamical interactions with other stars \citep[e.g.][]{wei+04}.
There are hints that the resonant torques by such a cusp can explain
some properties of the different sub-populations in the GC \citep{hop+06a},
however the direct detection of these torques requires test stars
on much shorter period orbits than are available at this time \citep[e.g.][]{mer+10}. 

We explore here a newly relevant \textbf{} dynamical method to probe
the dark cusp in the GC: binary evaporation constraints. Relatively
loosely-bound binaries are gradually detached by interactions with
the field stars around them; for an assumed cusp model this then imposes
limits on the properties of the surviving binary population \citep[e.g.][]{per09,hop09}.
Here we invert the constraints, and use the existence of binaries
in the GC to place upper limits on the product of the total number
density and the second moment of the mass distribution, $\n\eMM$,
or equivalently, a lower limit on the relaxation time $t_{\mathrm{rlx}}$,
and thus constrain dynamical models of the GC. The challenge is to
formulate the constraints in a way that provides robust estimates
in spite of statistical nature of evaporation and the uncertainty
in the binary's formation mechanism and dynamical age. It should be
emphasized that although binaries can be destroyed by processes other
than evaporation, this will not affect our estimates, which draw conclusions
only from those binaries that do survive. These provide an upper limit
on the effectiveness of each of the competing destruction mechanisms
separately, evaporation included, and thereby allow constraining the
physical parameters on which these processes depend.

The rest of this paper is organized as follows. The binary evaporation
timescale and its relation to the 2-body energy relaxation timescale
around a MBH are summarized in Sec. \ref{s:timescales}. Theoretical
upper bounds and several specific stellar density models for the GC
are presented in Sec. \ref{s:GCdensity} to serve as benchmarks for
evaluating the usefulness of binary evaporation constraints. The weak
limits from the currently know massive GC binaries are estimated in
Sec. \ref{s:limits}, and the potential of lower-mass, longer-lived
binaries to rule out density models is explored and discussed in detail.
Finally, the prospects and difficulties of the binary evaporation
method, both observational and theoretical, are discussed and summarized
in Sec. \ref{s:discussion}.

\section{Evaporation and relaxation timescales}

\label{s:timescales}

\subsection{The binary evaporation timescale}

\label{ss:tevap}

In an isotropic, Keplerian power-law cusp near an MBH, where the stellar
space density falls as $n_{\star}\propto r^{-\alpha}$, the Jeans
equation implies that the 1D velocity dispersion is \citep[e.g.][]{ale99}
\begin{equation}
\s^{2}(r)=\vbh^{2}(r)/(1+\alpha)\,,\label{e:sig2}
\end{equation}
where $v_{\bullet}^{2}=G\Mbh/r$. In a single mass population $\alpha=7/4$
\citep{bah+76}, and the same holds for the most massive component
of a multi-mass population \citep{bah+77}, if it is the dynamically
dominant one%
\footnote{\label{fn:Mseg}Many stellar populations can be approximated by a
2-mass mixture of heavy and light masses (stellar BHs with mass $M_{H}\sim10\,\Mo$
and number density $n_{H}$ far from the MBH, versus all the rest
with mass $M_{L}\sim1\,\Mo$, and number density $n_{L}$). In that
case the relaxational parameter $\Delta=4/(3+M_{H}/M_{L})(n_{H}M_{H}^{2})/(n_{L}M_{L}^{2})$
determined whether the relaxed steady state is in the strong segregation
limit ($\Delta\ll$1 and $\alpha_{H}=11/4$) or the weak segregation
limit ($\Delta\gg1$ and $\alpha_{H}=7/4$) \citep{ale+09}. %
} \citep{ale+09,kes+09}. Unless stated otherwise, we adopt this value
below. This velocity dispersion is in good agreement (to within 10\%)
with the one derived empirically by \citet{tri+08} for the old stellar
population on the $r=0.01-0.5$ pc scale. Note that well within the
radius of influence of the MBH ($\sim2$ pc in the GC), where the
MBH dominates the potential, the stars are \emph{not} in equipartition,
as the velocity dispersion depends on the stellar mass only through
the mass-dependence of the logarithmic slope, $\alpha(M_{\star})$,
which reflects mass segregation, and in steady state can vary between
$\alpha\sim3/2-5/2$. This limits the relative variations of $\sigma_{\star}^{2}(\Ms)$
to less than $\pm15\%$ in $\sigma_{\star}^{2}$ across the entire
stellar mass range (typically 3 orders of magnitudes in mass) \citep{ale+01a}.
We therefore assume here that all stars, as well as the centers of
mass (COM) of binaries, have the velocity dispersion given by Eq.
(\ref{e:sig2}).

A binary of mass $\Mb=M_{1}+M_{2}$, mass ratio $q_{m}=M_{2}/M_{1}\sim O(1)$
and binary semi-major axis (sma) $\ab=-GM_{1}M_{2}/2\Eb$ is considered
``soft'' when its binding energy $|E_{12}|$ is less than the typical
kinetic energy of the background stars of mean mass $\left\langle M_{\star}\right\rangle $
\citep[e.g.][]{heg75}, 
\begin{eqnarray}
\!\!\!\!\! s\!\equiv\!\frac{\left|\Eb\right|}{\eM\s^{2}} & \!=\! & \frac{1+\alpha}{2}\frac{M_{1}M_{2}}{\Mbh\eM}\frac{r}{\ab}\!\simeq\!\frac{1}{8}\frac{\Mb}{\eM}\frac{\vb^{2}}{\s^{2}(r)}\!<\!1\,,\label{e:soft}
\end{eqnarray}
where $r$ is taken to be the typical radius of the binary's orbit
around the MBH, $\vb^{2}\equiv G\Mb/\ab$, and the last approximate
equality assumes $q_{m}=1$. The parameter $s$ is the softness parameter---the
smaller $s$ is, the softer (relatively less bound) the binary. Statistically,
interactions with field stars make soft binaries softer, until they
become unbound, while hard binaries become harder, until they merge
\citep{heg75}. The present-day value of the softness parameter, $s_{0}$,
can be estimated for an assumed stellar density model ($\alpha$ and
$\eM$) from the observationally deduced properties of the binary
($\Mb$, $\ab$ and $r$). 

The evaporation timescale for soft binaries, whose COM velocity dispersion
is parametrized as $\sb^{2}=\s^{2}/q_{\sigma}$ (in equipartition,
$q_{\sigma}=\Mb/\Ms$, but near a MBH, $q_{\sigma}\simeq1$), is generally%
\footnote{Generalizing the derivation of \citet[Eq. 7.173]{bin+08}, who assume
$M_{12}=2M_{\star}$ and equipartition ($q_{\sigma}=2$). Note that
\citet{bin+87} assumed $q_{\sigma}=1$. %
} 
\begin{eqnarray}
\tau_{\mathrm{evap}} & \!\equiv\! & \frac{\left|\Eb\right|}{\left\langle D(\Delta E)\right\rangle }\!=\!\frac{1}{8}\sqrt{\frac{1+q_{\sigma}}{2\pi q_{\sigma}}}\frac{\Mb\s(r)}{G\n\eMM\ab\log\Lb(r)}\nonumber \\
 & \!\simeq\! & 0.07\frac{\vb^{2}\s(r)}{G^{2}\n\eMM\log\Lb(r)}\,\,\,(\mathrm{for\,}q_{\sigma}=1)\,,\label{e:tevap}
\end{eqnarray}
where $\Eb$ is the \emph{present-time} orbital energy, $\left\langle D(\Delta E)\right\rangle $
is the mean (i.e. averaged over binary COM velocities) energy diffusion
coefficient due to interactions of the binary with field stars, $\n$
is the total stellar number density, and $\eMM=\n^{-1}\int\mathrm{d}\Ms(\mathrm{d}n_{\star}/\mathrm{d}\Ms)\Ms^{2}$
is the second moment of the mass-function. $\eMM$ is usually dominated
by the heaviest mass in the mass spectrum (likely the stellar-mass
BHs)%
\footnote{For example, in the mass-segregated GC model of \citet{ale+09}, $\n\eMM=1.24n_{bh}M_{bh}^{2}=3.55\times10^{7}\,\Mo^{2}\,\mathrm{pc^{-3}}$
at $r=0.1$ pc.%
}, so that $\n\eMM\gtrsim n_{bh}M_{bh}^{2}$. At fixed binary mass,
the evaporation timescale scales approximately as $\tau_{\mathrm{evap}}\propto\eMM^{-1}$. 

The Coulomb factor for evaporation in the soft limit, $\Lb=b_{\max}/b_{\min}$,
is estimated by setting the maximal impact parameter to $b_{\max}=\ab/2$,
so that the interaction will be with only one of the binary members,
and not with the binary's COM, and the minimal one at the strong deflection
limit $b_{\min}=G(M_{1}+\Ms)/\left\langle v_{\mathrm{rel}}^{2}\right\rangle $,
where the typical relative velocity for an an encounter with a member
star of a soft binary is $\left\langle v_{\mathrm{rel}}^{2}\right\rangle =3\s^{2}+3\sb^{2}=3(1+1/q_{\sigma})\s^{2}$.
Therefore, 
\begin{eqnarray}
\Lb & \!\sim\! & 3\left(\frac{1\!+\!1/q_{\sigma}}{1\!+\!2/q_{\sigma}}\right)\!\frac{\s^{2}(r)}{\vb^{2}}\!=\!2\frac{\s^{2}(r)}{\vb^{2}}\!\simeq\!\frac{1}{4}\frac{M_{12}}{\eM}\frac{1}{s(r)}\,,\label{e:L12}
\end{eqnarray}
where $q_{\sigma}=1$ is assumed, and the last approximate equality
also assumes $q_{m}=1$.

Note that since binary evaporation requires that the perturber's distance
of closest approach be closer to one star than to the binary's center
of mass, the contribution of extended massive perturbers, such as
giant molecular clouds, to binary evaporation is substantially reduced
(see discussion in Sec. \ref{s:discussion}).

The \emph{timescale} $\tau_{\mathrm{evap}}$ is the present-time evaporation
timescale evaluated using the observed sma. Since $\left\langle D(\Delta E)\right\rangle $
is independent of $\Eb$ (neglecting logarithmic terms), it follows
that the \emph{actual time} to evaporation, $t_{\mathrm{evap}}$,
is 
\begin{equation}
\tev\simeq\int_{0}^{\Eb^{0}}\frac{\mathrm{d}E}{\left\langle D(\Delta E)\right\rangle }\simeq\frac{\Eb^{0}}{\left\langle D(\Delta E)\right\rangle _{t}}=\tau_{\mathrm{evap}}\frac{\Eb^{0}}{\Eb}\,,
\end{equation}
where $\Eb^{0}$ is the initial, unknown, binary energy and $\left\langle D(\Delta E)\right\rangle _{t}$
is the time-weighted average of the diffusion coefficient over the
binary's history%
\footnote{The binary's birth environment was likely much denser than its present
day environment. However a red giant binary spends a negligible fraction
of its lifetime there, and the uncertainty in the effective initial
softness (the one it has after leaving its birth environment) is absorbed
in the uncertainty in $\Eb^{0}$ (Eq. \ref{e:max-tevap}), and leads
to a weaker lower bound on $\tr$. %
}. Because the binary's evaporation is typically the result of many
weak interactions, the scatter of the actual time to evaporation around
the evaporation timescale is small, $\Delta\tev/\left\langle \tev\right\rangle \ll1$
\citep{per+07}. 

Since soft binaries become softer, a maximal bound on $\tev$ can
be obtained by assuming that the binary was initially as hard as it
could be, with $s=s_{\mathrm{hard}}$, which corresponded to the smaller
of either the soft/hard boundary $s=1$, or the hardest softness parameter
possible, that of a contact binary with an initial sma $\ab^{0}=\Rms{_{,1}}+\Rms{_{,2}}$,
where $\Rms$ is the radius of the star when still on the main sequence.
We therefore set 
\begin{equation}
s_{\mathrm{hard}}=\min[1,\mbox{s(}r;a_{12}=\Rms{_{,1}}+\Rms{_{,2}})]\,.\label{e:Sh}
\end{equation}
Since $s(t)\propto\Eb(t)$ (Eq. \ref{e:soft}), it follows that 
\begin{equation}
\tev\le\tau_{\mathrm{evap}}(s_{\mathrm{hard}}/s_{0})\equiv\tau_{\mathrm{evap}}S_{h}\,,
\end{equation}
where the maximal evolution ratio $S_{h}>1$ is the ratio of the maximal
initial and present-day softness parameters. For low-mass binaries,
where $\Mb\sim O(2\,\Mo)$ and $\Rms\sim O(1\,\Ro)$, $s_{\mathrm{hard}}\sim0.5\left(\eM/1\,\Mo\right)^{-1}(r/0.1\,\mathrm{pc})$.
We adopt below 
\begin{equation}
\tev=\tau_{evap}S_{h}\,,\label{e:max-tevap}
\end{equation}
 as a conservative over-estimate of the evaporation time, which leads
to a conservative under-estimate of the lower bound on the relaxation
time (see Eq. \ref{e:mintr}).

\subsection{The 2-body relaxation timescale}

\label{ss:trlx}

The local Chandrasekhar 2-body energy relaxation time \citep[e.g.][]{spi87}
is 
\begin{eqnarray}
\tr & = & 0.34\frac{\s^{3}(r)}{G^{2}\n\eMM\log\Lambda}\label{e:trlx}\\
 & \simeq & \frac{0.68}{(3-\alpha)(1+\alpha)^{3/2}}\frac{\Mbh^{2}}{\eMM}\frac{P(r)}{N_{\star}(r)\log\Lambda}\,,\nonumber 
\end{eqnarray}
where $N_{\star}(r)$ is the number of stars enclosed within $r$,
and $\Lambda\sim\Mbh/\Ms$ \citep{bar+13}. Note that $\Lambda\gg\Lambda_{12}$,
since in contrast to evaporation, encounters on all scales contribute
to relaxation. It therefore follows that
\begin{equation}
\tr\simeq4.8\frac{\log\Lb(r)}{\log\Lambda}\frac{\s^{2}(r)}{\vb^{2}}\tau_{\mathrm{evap}}\gg\,\tau_{\mathrm{evap}}.\label{e:trlx-tevap}
\end{equation}
Thus a lower limit on the evaporation timescale translates to a lower
limit on the 2-body relaxation time, and an upper limit on $\n\eMM$.
An observed binary of period $\Pb$, mass $\Mb$ and age $\Tb\le\tev\le\tau_{\mathrm{evap}}S_{h}$
at distance $r$ from the MBH implies therefore that the 2-body relaxation
timescale there exceeds 

\begin{equation}
\mtr(r)\simeq1.4\frac{\log\Lb(r)}{\log\Lambda}\frac{\s^{2}(r)}{\vb^{2}}\frac{\Tb}{S_{h}(r)}\,.\label{e:mintr}
\end{equation}

Note that the estimator $\mtr$ does not depend on $\eMM$, as $\tr$
does, but rather on the less sensitive $\eM$, which enters weakly
through the Coulomb logarithm%
\footnote{The exact dependence of the Coulomb logarithm on the mass function
has not yet been studied in detail, to the best of our knowledge.%
} $\log\Lambda=\Mbh/\eM$, and more strongly through the factor $S_{h}$.
All the other quantities that appear in $\mtr$ are directly observable
or derivable: $\log\Lambda_{12}$ (Eq. \ref{e:L12}), $\sigma_{\star}$
(Eq. \ref{e:sig2}) and $T_{12}$. This lower bound can be used to
rule out a specific density model for the GC, which predicts $\alpha$
and $\eM$. Alternatively, the uncertainty in $\mtr$ can be estimated
by noting that when $s_{\mathrm{hard}}<1$, $\mtr\propto\Rms(\Ms)/\log(\Mbh/\Ms)$,
where $R_{MS}(\Ms)\propto\Ms^{\beta}$, with $\beta=1.25$ below $\sim1.5\,\Mo$,
at the transition between p-p to CNO cycle hydrogen burning, to $\beta=0.56$
above it \citep{sch+92a}. The maximal range of $\eM$, which lies
between $\sim1\,\Mo$ (MS dwarfs) and $\sim10\,\Mo$ (stellar BHs),
then implies the range in $\Rms$ of $\sim1\,\Ro$ to $\Rms\lesssim10\,\Ro$
(for an $O(100\,\Mo)$ BH progenitor), which translates to one order
of magnitude uncertainty in $\mtr$. When $s_{\mathrm{hard}}=1$,
$\mtr\propto1/\eM\log(\Mbh/\Ms)$.

The lower bound on the relaxation time translates into an upper bound
on $\n\eMM$ (Eqs. \ref{e:trlx}, \ref{e:mintr}),

\begin{equation}
\mnMM\simeq0.24S_{h}(r)\frac{\s(r)\vb^{2}}{G^{2}\log\Lb(r)T_{12}}\,.\label{e:maxnM2}
\end{equation}
The estimate $\mnMM$ does not itself depend on $\eMM$. An upper
bound on the total stellar number density $\n$ can be obtained for
an assumed value of $\eMM$.

\subsection{Binary formation by 3-body captures}

\label{ss:3body}

The use of long-period binaries to constrain the dark cusp density
requires reliable estimates of their \emph{dynamical} age. For a primordial
binary, the dynamical age equals the stellar age, which can be estimated
from the binary's luminosity and spectrum. However, when the binary
is formed dynamically long after the stars are born, the dynamical
age can only be bound from above by the stellar age. Such binaries
can not provide strong constraints on the cusp density. Unless non-primordial
binaries can be identified, their existence may be misinterpreted
and lead to erroneous conclusions. However, as we argue here, such
locally-formed binaries are expected to be extremely rare near a MBH.
Another source of contamination, one that can not be decisively ruled
out, is tidal capture of hierarchical triples in 4-body interactions
with the MBH. This is discussed separately in Sec. \ref{s:discussion}. 

Of the mechanisms generally considered for binary formation \citep[e.g.][]{toh02},
those decoupled from the formation of the stars themselves are tidal
capture in 2-body interactions and dynamical capture by 3-body interactions.
Tidal capture is generally an inefficient process, leading to mergers
rather than binary formation \citep[see e.g. review by][]{ben06};
it is very inefficient in galactic nuclei, since stellar encounters
in the deep potential of a MBH are strongly hyperbolic---close encounters
lead to mass-loss and spin-up, rather than capture \citep{ale+01a}.
Binary formation by 3-body capture requires that three stars to interact
with mutual impact parameters of the order of $b_{\pi/2}\sim GM_{\star}/\sigma^{2}$
\citep[Eq. 7.10]{bin+08}. This then allows strong interactions ($\Delta v\sim v$)
that extract a significant amount of relative kinetic energy from
one pair and enable it to form a bound binary. The sma of the newly
formed binary is expected to be also of order $a_{12}\sim b_{\pi/2}$.
 It then follows that a typical binary formed by 3-body capture is
soft, with an initial softness parameters of $s_{0}\sim M_{12}/4\eM\sim1/2<1$
(Eq. \ref{e:soft} with $M_{12}=2\eM$), as also suggested by the
detailed analysis of 3-body capture in globular clusters \citep{goo+93}.
The formation rate is then $\sim N_{\star}(r)/t_{3}$, where $t_{3}$
is the typical timescale for a star to be involved in a 3-body capture
interaction \citep[Eq. 7.11]{bin+08}, 

\begin{equation}
t_{3}\sim\frac{\s^{9}}{n_{\star}^{2}G^{5}M_{\star}^{5}}\,.
\end{equation}
The number of binaries formed by 3-body encounters on the spatial
scale $r$ over a time $t$ a relaxation time (Eq. \ref{e:trlx})
is therefore 
\begin{equation}
N_{12}(r;t)\!\sim\! N_{\star}(r)\frac{t}{t_{3}}\!\sim\! N_{\star}^{2}(r)\left(\frac{\Ms}{\Mbh}\right)^{3}\frac{(1\!+\!\alpha)^{3}}{\log\Lambda}\frac{t}{t_{\mathrm{rlx}}}\,.
\end{equation}
 After a Hubble time, $t_{H}=10\,\mathrm{Gyr}$, $N_{12}(t_{H})\sim10^{-8}(t_{H}/t_{\mathrm{rlx}})\lesssim10^{-6}$
for values typical of the GC ($\Mbh/\eM\sim10^{6}$ and $N_{\star}(r\!=\!0.1\,\mathrm{pc)\sim10^{5}}$),
and for the shortest realistic value for $t_{\mathrm{rlx}}$, $\min t_{\mathrm{rlx}}(0.1\,\mathrm{pc})>10^{8}\,\mathrm{yr}$
(see Sec. \ref{ss:drainlim}). The contribution of this formation
channel to the binary population in the GC is therefore negligible
\citep[see also][]{hop09}.

\section{Stellar density models for the Galactic center}

\label{s:GCdensity}

To evaluate the practical relevance of GC binaries for constraining
the dark cusp density, it is useful to have benchmarks for possible
density profiles. We consider here a general dynamical upper bound
on the cusp density, and two classes of specific models, one with
a relaxed, high-density mass-segregated cusp, and the other with an
unrelaxed low-density core.

\subsection{Dynamical bounds on the dark cusp}

\label{ss:drainlim}

Mass segregation models of the GC, which assume a passive evolution
of the nuclear cluster, predict the accumulation of a steady state
high density cusp of stellar remnants \citep{mor93}. In steady state,
the rate at which stars are scattered into the MBH from some volume
$<r$ is balanced by the rate at which stars are replenished by 2-body
relaxation from outside $r$. The detailed calculation of the steady
state density can be robustly bracketed by the conservative ``drain
limit'' \citep{ale+04}, which estimates the maximal number of objects
that can survive against mutual 2-body scattering into the MBH over
the age of the system. Taking the age of the system to be the Hubble
time, the drain limit is the solution $N_{\star}(t=0)$ of the condition
\begin{equation}
\frac{1}{N_{\star}}\frac{\mathrm{d}N_{\star}}{\mathrm{d}t}=\frac{1}{\log(\sqrt{2r/r_{g}})\tr(r)}<\frac{1}{t_{H}}\,,
\end{equation}
where $r_{g}=G\Mbh/c^{2}$ is the gravitational radius of the MBH.
The drain limit corresponds to a minimal required survival probability
of $1/2$ over a Hubble time, and it is only a function of $\eMM$
(up to a logarithmic term%
\footnote{Approximating $t_{\mathrm{rlx}}\sim\sigma^{3}/G^{2}M_{\star}^{2}n_{\star}\log\Lambda$,
$\sigma^{2}\sim GM_{\bullet}/r$ and $N_{\star}\sim(4\pi/3)n_{\star}r^{3}$.%
}), 

\begin{equation}
\max N_{\star}(<r)\sim\frac{2}{3}\frac{\log(\sqrt{2r/r_{g}})}{\log(\Mbh/\eM)}\frac{\Mbh^{2}}{\eMM}\frac{P(r)}{t_{H}}\,,\label{e:drain}
\end{equation}
where $P(r)$ is the orbital period around the MBH. It then follows
that the shortest possible relaxation time in steady state, which
is attained at the maximal density of the drain limit (substituting
$\max N_{\star}$ into Eq. \ref{e:trlx}), does not depend on the
assumed mass function,

\begin{equation}
\mtr{_{,\mathrm{drain}}}\simeq\frac{1}{(3-\alpha)(1+\alpha)^{3/2}}\frac{1}{\log(\sqrt{2r/r_{g}})}t_{H}\,,\label{e:mintrDL}
\end{equation}
For $\alpha=1.75$ this corresponds to 
\begin{equation}
\mtr{_{,\mathrm{drain}}}\simeq2.5\times10^{8}\,\mathrm{yr}\,.
\end{equation}
Varying the assumed $\alpha$ between $\alpha=0\to2$ changes the
minimal relaxation time between $2.4\times10^{8}\,\mathrm{yr}$ (for
$\alpha=7/5=1.4$) to a maximal one of $4.7\times10^{8}\,\mathrm{yr}$
(for $\alpha=0$). 

The drain limit is a useful benchmark for our purpose since it provides
an estimate of the shortest physically realizable relaxation time
in steady state, which does not depend on detailed features of the
Galactic model, such as the mass function and the boundary conditions.
For example, the upper bounds on the enclosed number and density on
the scale of the orbit of IRS 16NE, at $r=0.15\,\mathrm{pc}$ (Sec.
\ref{ss:IRS16NE}), are 
\begin{eqnarray}
\max N_{\star}\! & \sim\! & 1.8\!\times\!10^{4}\frac{(10\,\Mo)^{2}}{\eMM}\,,\nonumber \\
\max\n & \!\sim\! & 1.25\!\times\!10^{6}\frac{(10\,\Mo)^{2}}{\eMM}\!\left(\frac{3\!-\!\alpha}{3}\right)\,\mathrm{pc^{-3}}\,.
\end{eqnarray}
The drain limit is only a few times higher than the values derived
by detailed calculations. For example, inside the inner $\sim O(0.01\,\mathrm{pc)}$,
the mass density of the GC is expected to be dominated by stellar
mass BHs with mass $M_{bh}\sim O(10\,\Mo)$ (Figure \ref{f:meanM})
\citep{fre+06,ale+09}, and in that case $\eMM^{1/2}\sim M_{bh}$
and $\max N_{\star}\sim\max N_{bh}$. The drain limit on $10\,\Mo$
stellar BHs in the inner $0.01$ pc is $\max N_{bh}(<0.01\,\mathrm{pc)\simeq250}$,
compared to $N_{bh}(<0.01\,\mathrm{pc)\simeq100}$ derived by Fokker-Planck
models, whether solved numerically (\citealt{ale+09}; see also Figure
\ref{f:IMFs}), or by the H\'enon (Monte-Carlo) method \citep{fre+06}.

\subsection{A relaxed mass-segregated high density cusp }

The evolution of a galactic nucleus to a relaxed steady state configuration
can be approximated in terms of the evolution of the stellar distribution
function (DF) due to diffusion in phase space, and integrated in time
from an arbitrary initial configuration to steady state by the Fokker-Planck
(FP) equation. This scheme was first applied to a multi-mass system
with a central MBH by \citet{bah+77}, with the approximations that
the potential is Keplerian, the angular momentum diffusion is averaged,
so that only diffusion in energy is integrated in time, and only non-coherent
2-body relaxation is considered. Two boundary conditions are imposed,
one at very high energy (close to the MBH), where the DF falls to
zero, and one at the energy corresponding to the MBH's radius of influence
($r_{h}\sim2\,\mathrm{pc}$ in the GC), where the stellar number density
of each mass is fixed to some value, which reflects its population
ratio far from the dynamical influences of the MBH, and therefore
depends only on the assumed initial mass function (IMF) and the star-formation
history.

\begin{figure}
\noindent \begin{centering}
\begin{tabular}{c}
\includegraphics[width=1\columnwidth]{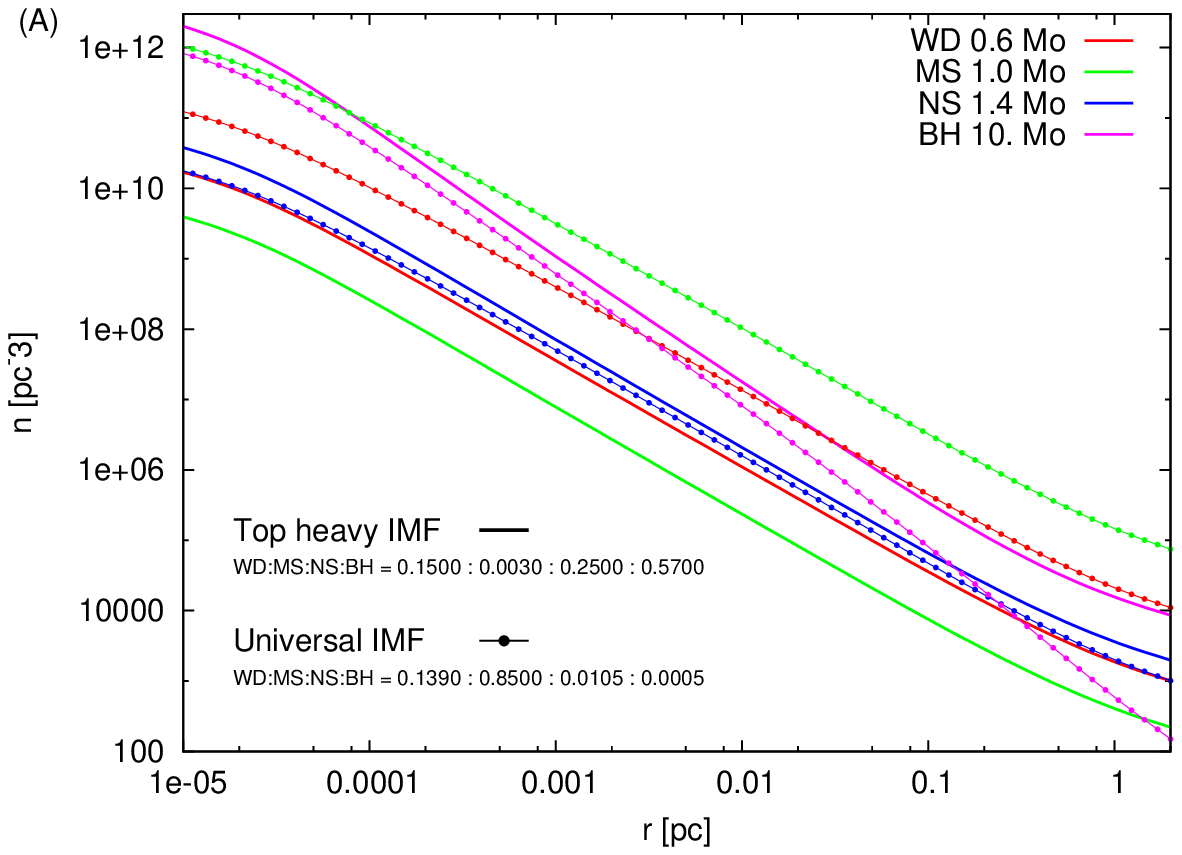}\tabularnewline
\includegraphics[width=1\columnwidth]{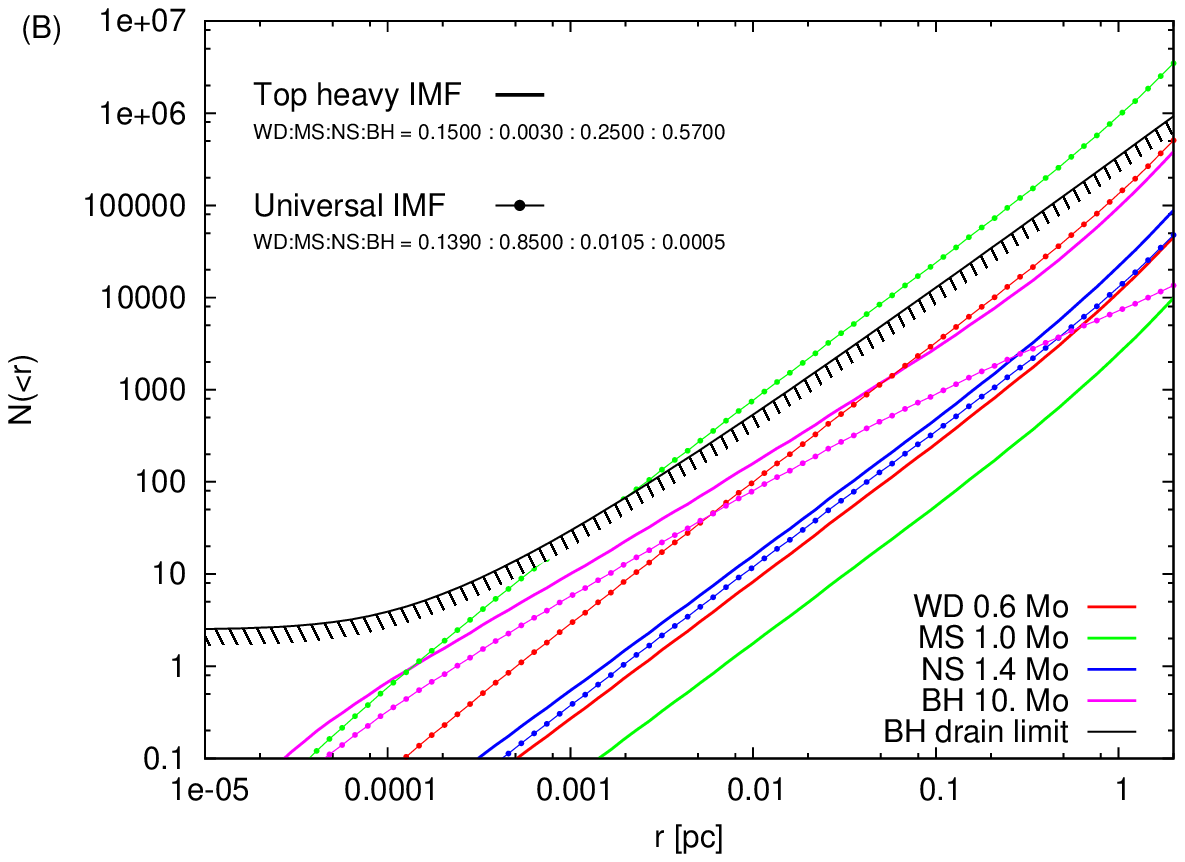}\tabularnewline
\end{tabular}
\par\end{centering}

\caption{\label{f:IMFs}(A) Fokker-Planck steady-state mass-segregated density
models for the GC \citep[cf][]{ale+09}, assuming a simplified 4-component
mass function ($0.6\,\Mo$ white dwarfs, $1\,\Mo$ main-sequence stars,
$1.4\,\Mo$ neutron stars and $10\,\Mo$ stellar BHs). One model has
the ``universal'' IMF (modeled here by the \citet{mil+79} IMF),
and the other has an extreme top-heavy IMF with $\mathrm{d}\n/\mathrm{d}\Ms\propto\Ms^{-0.45}$
between $\Ms=0.8\, M_{\odot}$ to $120\,\Mo$. The number ratios of
the four components in each of the models are listed on the plot (see
also table \ref{t:PMF}). (B) The cumulative number of stars in the
universal and top-heavy GC models, are compared to each other and
to the drain limit for stellar BHs, which both models obey. }
\end{figure}

Figure (\ref{f:IMFs}) shows results from such simple FP models for
the GC, which assume present-day mass functions (PMFs) of stars and
compact remnants that are derived by stellar population synthesis
\citep{ste98} for continuous star-formation scenario for the GC \citep{ale+99a,fig+04},
assuming two different IMFs (table \ref{t:PMF}). The first is based
on the ``universal'' IMF (here approximated by the \citealt{mil+79}
IMF, extending between $\Ms=0.1\, M_{\odot}$ to $120\,\Mo$), which
reproduces the volume-averaged $K$-band luminosity of the luminous
red giants in the inner few pc of the GC \citep{ale+99a} and a second,
extremely top-heavy model that is suggested by the population of the
massive disk stars \citep{bar+10}, with $\mathrm{d}\n/\mathrm{d}\Ms\propto\Ms^{-0.45}$
between $\Ms=0.8\, M_{\odot}$ to $120\,\Mo$. The two IMFs result
in very different stellar populations. While the mass and number of
the universal IMF model are dominated by main sequence stars, and
the BHs are only $\sim0.01$ of the mass and $5\times10^{-4}$ of
the number, the top-heavy IMF model is completely dominated by the
compact remnants, with the BHs $0.89$ of the mass and $0.57$ of
the number. Nevertheless, the relaxed steady states of both models
conform with the drain limit (Figure \ref{f:IMFs}B), which demonstrates
its robustness. 

A comparison of the density profiles of the BHs outside of the inner
$\sim0.01\,\mathrm{pc}$ in the two IMF models (Figure \ref{f:IMFs}A)
reveals the difference between the weak and strong mass-segregation
regimes \citep{ale+09}. When the BHs, the most massive long-lived
component in the mass function, are dominant in the population as
they are in the top-heavy model%
\footnote{{\footnotesize{The top-heavy model has $\Delta\simeq40\gg1$ (weak
segregation limit), whereas the universal model has $\Delta\simeq0.4<1$
(strong segregation limit), counting all stars and remnants with mass
$<2\,\Mo$ as light (see footnote \ref{fn:Mseg}).}}%
}, they relax to an $\alpha=7/4$ cusp ($\n\propto r^{-\alpha}$);
when they are not, as in the universal model, they relax to a steeper
slope, with $\alpha\gtrsim2$. Closer to the MBH (here, at $r\lesssim0.01$
pc), mass segregation makes the BHs locally dominant, and both models
converge to the $\alpha=7/4$ configuration. 

\begin{figure}
\noindent \begin{centering}
\begin{tabular}{c}
\includegraphics[width=1\columnwidth]{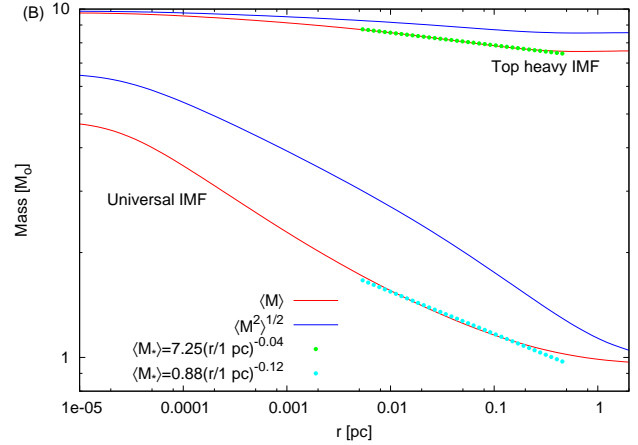}\tabularnewline
\end{tabular}
\par\end{centering}

\caption{\label{f:meanM} The first and second moments of the stellar mass,
$\eM$ and $\eMM$, as function of distance from the MBH in the two
FP models for the GC (Figure \ref{f:meanM}). Also shown is a simple
analytic approximation for $\eM$ in the range $r=0.005-0.5$ pc. }
\end{figure}

Figure (\ref{f:meanM}) shows the variation of the the first and second
mass moments ($\eM$ and $\eMM$) as function of distance from the
MBH, for the two models, and provides convenient power-law approximations
for $\eM$, for use in the minimal relaxation time estimate, Eq. (\ref{e:mintr}).

\begin{table*}
\caption{\label{t:PMF}GC stellar PMF models (unsegregated)}

\begin{centering}
\begin{tabular}{lc|cc|cc|l}
\hline 
 & \multicolumn{1}{c}{} & \multicolumn{2}{c}{Universal IMF} & \multicolumn{2}{c}{Top-heavy IMF} & \tabularnewline
 &  & Mass  & Number & Mass & Number & \tabularnewline
Stellar type & Typical mass & fraction & fraction & fraction & fraction & Comments\tabularnewline
\hline 
\hline 
Main sequence & $0.43\,\Mo/7.22\,\Mo$ $^{1}$ & $0.73$ & $0.85$ & $0.03$ & $0.03$ & Mean mass incl. remnants $0.48\,\Mo/6.38\,\Mo$ $^{1}$\tabularnewline
White dwarfs & $0.6-1.1\,\Mo$ & $0.23$ & $0.14$ & $0.02$  & $0.14$ & Progenitor mass $2.5-8\,\Mo$\tabularnewline
Neutron stars & $1.4\,\Mo$ & $0.03$ & $0.01$ & $0.06$ & $0.26$ & Progenitor mass $8-30\,\Mo$\tabularnewline
Black holes & $10\,\Mo$ & $0.01$ & $0.0005$ & $0.89$ & $0.57$ & Progenitor mass $>30\,\Mo$\tabularnewline
\hline 
\hline 
\multicolumn{7}{l}{{\footnotesize{$^{1}$ Average mass for the universal IMF and the
top-heavy IMF, respectively}}.}\tabularnewline
\hline 
\end{tabular}
\par\end{centering}

\end{table*}

\subsection{An unrelaxed low-density Galactic core}

The depleted numbers of red giants in the GC could be the result of
a past merger event with another MBH, which dynamically ejected stars
from the center \citep{mer10}. The resulting drastic decrease in
the stellar density at the center would then increase the 2-body relaxation
time well above the Hubble time, thereby slowing down the the re-formation
of a relaxed cusp. For certain choices of the initial radius of the
evacuated region and of the time before present when the merger occurred,
the slowly evolving post-merger GC density profile can match that
observed today. Figure (\ref{f:min_trlx_r}) shows the resulting 2-body
relaxation time \citet{mer10}, which substantially exceeds the Hubble
time in the inner $\sim0.2$ pc of the GC.

\section{Binary evaporation limits on the dark cusp in the GC}

\label{s:limits}

\subsection{Long-period young massive binaries }

\label{ss:IRS16NE}

The recently discovered long-period massive binary IRS 16NE \citep{pfu+13}
at a projected distance of $p=3.05^{\prime\prime}$ \citep{pau+06}
has a period of $P=224.09\pm0.09$ days and a mass in the range $80\,\Mo\lesssim\Mb\lesssim100\,\Mo$,
based on the similarity of the observed spectrum of the primary to
the known eclipsing binary IRS 16SW with $M_{1}\sim50\,\Mo$ \citep{mar+06b,mar+07}.
The absence of spectral evidence for the secondary further suggests
that $M_{2}\lesssim30\,\Mo$, so that the binary mass likely lies
toward the lower limit of this mass range. IRS 16NE is a member of
the clock-wise disk, a product of the recent star-burst episode, whose
age is estimated at $T=(6\pm2)\times10^{6}\,\mathrm{yr}$ (\citealt{pau+06,bar+09};
but see also younger age estimate by \citealt{lu+13}). 

We therefore adopt below a binary age of $\Tb=6\times10^{6}\,\mathrm{yr}$
and binary mass of $\Mb=80\,\Mo$, which translate to an sma $\ab=3.11\,\mathrm{AU}$
and $\vb=151\,\mathrm{km\, s^{-1}}$ (for $M_{12}=100\,\Mo$, $\ab=3.26\,\mathrm{AU}$).
The 3D distance of IRS 16NE and its eccentricity can be estimated
using the inferred distance to the GC, $R_{0}\simeq8.3\,\mathrm{kpc}$
\citep{gil+09_short}, the binary's observed plane-of-sky position
and 3 components of the velocity, and by assuming that it is orbiting
in the plane of the counter clockwise disk \citet{pau+06}. It then
follows that $a\mathrm{\simeq0.15}\,\mathrm{pc}$%
\footnote{A similar value is derived by assuming spherical symmetry, and estimating
a 3D distance of $r=\sqrt{3/2}p0.04\,\mathrm{pc\simeq0.15\,}\mathrm{pc}$.%
} and $e\simeq0.14$ ($r\simeq0.13$ pc, $v_{\bullet}\simeq350\,\mathrm{km\, s^{-1}}$).
 Since the binary is on a nearly circular orbit, it is justified
to substitute $r=a$ and calculate the relevant timescales without
orbital averaging. We further assume, based on the mass-segregated
steady state model of the GC \citep{ale+09}, that the mean logarithmic
slope at $r=0.15$ pc is $\alpha=1.5$, and that $\eM=1.2\,\Mo$ and
$\eMM^{1/2}=1.8\,\Mo$. The Coulomb logarithm for evaporation is then
$\log\Lb\simeq1.46$. According to the binary softness criterion (Eq.
\ref{e:soft}), IRS 16NE is soft, or marginally soft, with $s_{0}$
in the range $0.2$ (for $M_{12}=80\,\Mo$ and $\alpha=0$) to $0.9$
(for $M_{12}=100\,\Mo$ and $\alpha=2$). We therefore assume that
the soft binary evaporation timescale is relevant for estimating the
relaxation time scale. Since $\Rms(50\,\Mo)=9.5\,\Ro$ and $\Rms(30\,\Mo)=7.1\,\Ro$,
$s_{\mathrm{hard}}=1$, and therefore the existence of IRS 16NE implies
\begin{equation}
\mtr(\sim0.15\,\mathrm{pc)\sim2\times10^{7}}\,\mathrm{yr}\,.
\end{equation}
The corresponding upper limit on the total number density 
\begin{equation}
\left.\max\n=\mnMM\right/\eMM\sim2\times10^{8}\,\mathrm{pc^{-3}}\,.
\end{equation}

Varying the assumed parameters over the range $\Tb=4\to8\,\mathrm{Myr}$,
$\Mb=50\to100\,\Mo$, $\alpha=0\to2$ changes $\mtr$ between $\sim1\times10^{7}\,\mathrm{yr}$
(for $\alpha=2$, $\Mb=50\,\Mo$ and $\Tb=4\times10^{6}\,\mathrm{yr}$)
to $\sim6\times10^{7}\,\mathrm{yr}$ (for $\alpha=0$, $\Mb=100\,\Mo$
and $\Tb=8\times10^{6}\,\mathrm{yr}$). Such a short lower bound on
the relaxation timescale lies well below the drain limit (Figure \ref{f:min_trlx_a}
B), and therefore cannot constrain realistic density models (see also
\citealt{ant+11}). This conclusion is reinforced if instead $\eM=10\,\Mo$
is assumed, which further reduces $\mtr$ by a factor $O(10)$.

\subsection{Long-period old low-mass binaries}

\label{sss:RG}

Low-mass binaries have not yet been detected in the GC. Estimates
of their numbers there are highly uncertain, given that star formation
and dynamics in the unusual dense environment near a MBH are not well
understood. The statistics of dense Galactic clusters could provide
relevant estimates, as hinted by the finding of \citet{pfu+13} that
the number fraction of massive binaries in the GC ($F_{2}=0.08$--$0.56$
at the 0.95 confidence level), which is consistent with that observed
in dense Galactic OB clusters. Stellar clusters can provide relevant
guidelines for low-mass binaries. \citet{sol+07} find in a sample
of low-density Galactic globular clusters a low-mass binary fraction
$F_{2}>0.06$ in the cluster core, and globally $F_{2}\sim0.1$--$0.5$
over the entire cluster. \citet{som+09} find a similar $F_{2}>0.05$
lower limit on the binary fraction in the core, and $F_{2}>0.03$
outside of it in globular cluster M4. Of these binaries, only a fraction
can be actually detected. We now discuss the observational constraints,
the red giant population in the GC, and the dynamical information
that can be derived from the detection of long-period low mass binaries
in the GC.

\subsubsection{Detection of long-period binaries in the GC }

In order for a binary to be both detectable and useful for probing
the density of the hypothetical dark cusp, it must satisfy several
constraints. 

(1) The binary's orbital period $P_{12}$ should be short enough relative
to the monitoring period, and the observing cadence high enough, so
as to allow the extraction of the orbital parameters from the data,
\begin{eqnarray}
\max\Pb=\min2\pi\sqrt{\ab^{3}/G\Mb} & = & x_{P}T_{\mathrm{obs}}\,,\label{e:Pdetect}
\end{eqnarray}
where the factor $x_{P}$ depends on the detection mode. For spectroscopic
binaries $x_{P}\sim1/2$. However, for visual binaries a value as
large as $x_{P}\sim10$ may be feasible, as discussed further below.

(2) The binary should be luminous enough to be observed with high
S/N for reliable determination of the reflex velocity (if observed
as a spectroscopic binary) or proper motion (if observed as a visual
binary), with a magnitude of 
\begin{equation}
\max K<K_{\mathrm{S/N}}\,,\label{e:Kdetect}
\end{equation}
where $K$ is the magnitude of a single star.

(3) The binary's reflex motion or proper motion should be large enough
to be measured with high S/N, 

\begin{equation}
\min\vb=\min\sqrt{G\Mb/\ab}>v_{\mathrm{S/N}}\,.\label{e:vdetect}
\end{equation}

(4) The binary should be dynamically sensitive to the relaxation timescales
of interest, and at least to the conservative drain limit (Eq. \ref{f:meanM}).
This would be the case if it is long-lived, with a main-sequence age
$T_{MS}$ such that (Eq. \ref{e:mintr}) 
\begin{equation}
\min T_{MS}={\cal O}(10)\frac{\vb^{2}}{\vbh^{2}(r)}S_{h}(r)\tr>{\cal O}(10)\frac{\vb^{2}}{\vbh^{2}(r)}\tr\,,\label{e:TMSdetect}
\end{equation}
where $S_{h}\ge1$ is the ratio between the initial and present day
softness parameter (Eq. \ref{e:Sh}). 

(5) Long-lived low-mass main-sequence stars in the GC, with $\Ms\lesssim2\,\Mo$
are faint, with $K_{MS}\gtrsim19$%
\footnote{For $\mathrm{DM}_{GC}=14.6$ \citep{gil+09_short} and $A_{K}=2.7$
\citep{fri+11}, taking $M_{K}(2\,\Mo)=1.6$ and $M_{K}(1\,\Mo)=3.75$
(adapted from \citet{sch+92a} ZAMS models for $Z=0.02$).%
}. This is still several magnitudes fainter than current limits for
reliable spectroscopic identification%
\footnote{The photometric detection limit in the GC is $K\sim19$ \citep{sch+07}.
Due to stellar crowding, the detection completeness starts to be affected
already at a magnitude of $K<17$ \citep{sch+07}. The faintest stars
that have been spectroscopically identified in the GC are $K\sim17.6$
\citep{pfu+11}, yet with an integration time of $\sim10$ hr and
low signal-to-noise. Typical integration times of 1-2 hr allow the
spectroscopic identification of stars with $K\sim14$--$15$ \citep[e.g.][]{bar+10}. %
}. Low-mass stars can only be observed in the giant phase, when their
size expands to $\Rs\sim\mathrm{few\times10\,\Ro}\sim{\cal O}(0.1\,\mathrm{AU})$.
This limitation decreases the available targets by a factor of $\sim10$
(the typical ratio of the time spent in the giant phase relative to
$T_{MS}$), and imposes an additional requirement on the size of the
orbit, $\Rs<\ab/2$, to avoid the hard-to-model complications of common
envelope evolution, or equivalently, 
\begin{equation}
\max v_{12}<\vs\,,\label{e:Rdetect}
\end{equation}
where $\vs=\sqrt{G\Ms/\Rs}$ is the red giant's circular velocity,
and $\Ms\simeq\Mb/2$ is assumed. For low-mass red giants, $\vs\sim{\cal O}(10^{2}\,\mathrm{km\, s^{-1}})$.
However, these limitations are mitigated by the fact that red giants
are advantageous for precise spectroscopic velocity determination
since they are slow rotators and have sharp molecular absorption lines.

We now consider specifically the constraints imposed by present-day
and upcoming observing capabilities of the VLT. It is straightforward
to scale these constraints to other instruments and future capabilities. 

Current observations of the GC with the SINFONI integral field spectrograph
on the VLT \citep{eis+03b,bon+04} reach a radial velocity uncertainty
of $\lesssim5\,\mathrm{km\, s^{-1}}$ in a 1 hr integration for a
$K<K_{S/N}\simeq14$ red giant. A $3\sigma$ detection of the reflex
motion then requires a radial velocity curve amplitude of $V_{r}\gtrsim15\,\mathrm{km\, s^{-1}}$.
Assuming a circular orbit and the mean decrease by a factor $\pi/4$
in the line-of-sight velocity due to the orbital inclination, $v_{\mathrm{S/N}}\simeq20\,\mathrm{km\, s^{-1}}$.
Since typically $v_{\mathrm{S/N}}<v_{\star}$, the red giant size
constraint (Eq. \ref{e:Rdetect}) is not a limitation. The minimal
useful main-sequence lifetime can be estimated by taking $\vb$ in
the range $v_{\mathrm{S/N}}$ to $\vs$ (Eq. \ref{e:TMSdetect}).
For example, for $r=0.15\,\mathrm{pc}$, where $\mtr{_{,\mathrm{drain}}}\simeq2.5\times10^{8}\,\mathrm{yr}$
(Sec. \ref{ss:drainlim}), $\min T_{MS}\sim8\times10^{6}-2\times10^{8}\,\mathrm{yr}$,
which corresponds to a maximal useful mass in the range $\max\Ms\sim4-20\,\Mo$
\citet{sch+92a}. More realistic dynamical models with $\tr>\mtr{_{,\mathrm{drain}}}$
(Sec. \ref{s:GCdensity}) limit $\max\Ms$ to the lower boundary of
this range.

\begin{figure}
\noindent \centering{}%
\begin{tabular}{c}
\includegraphics[width=1\columnwidth]{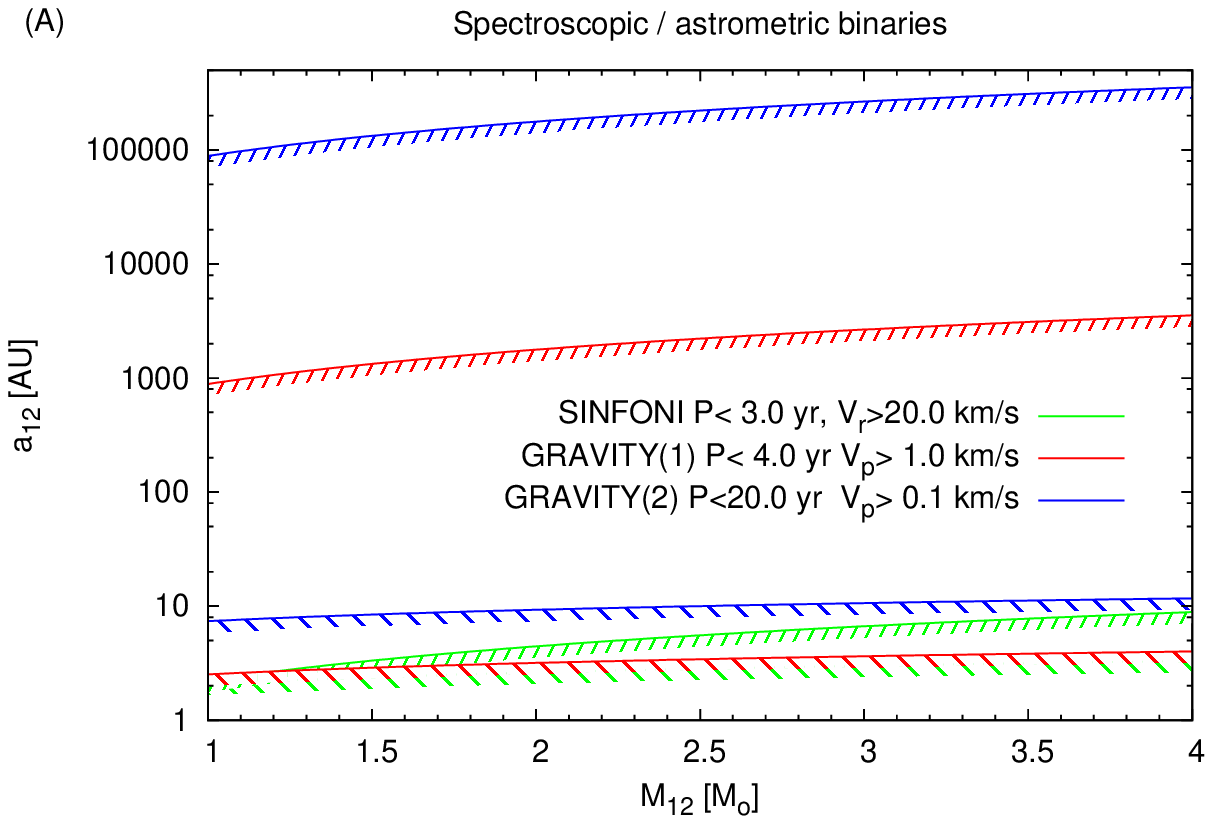}\tabularnewline
\includegraphics[width=1\columnwidth]{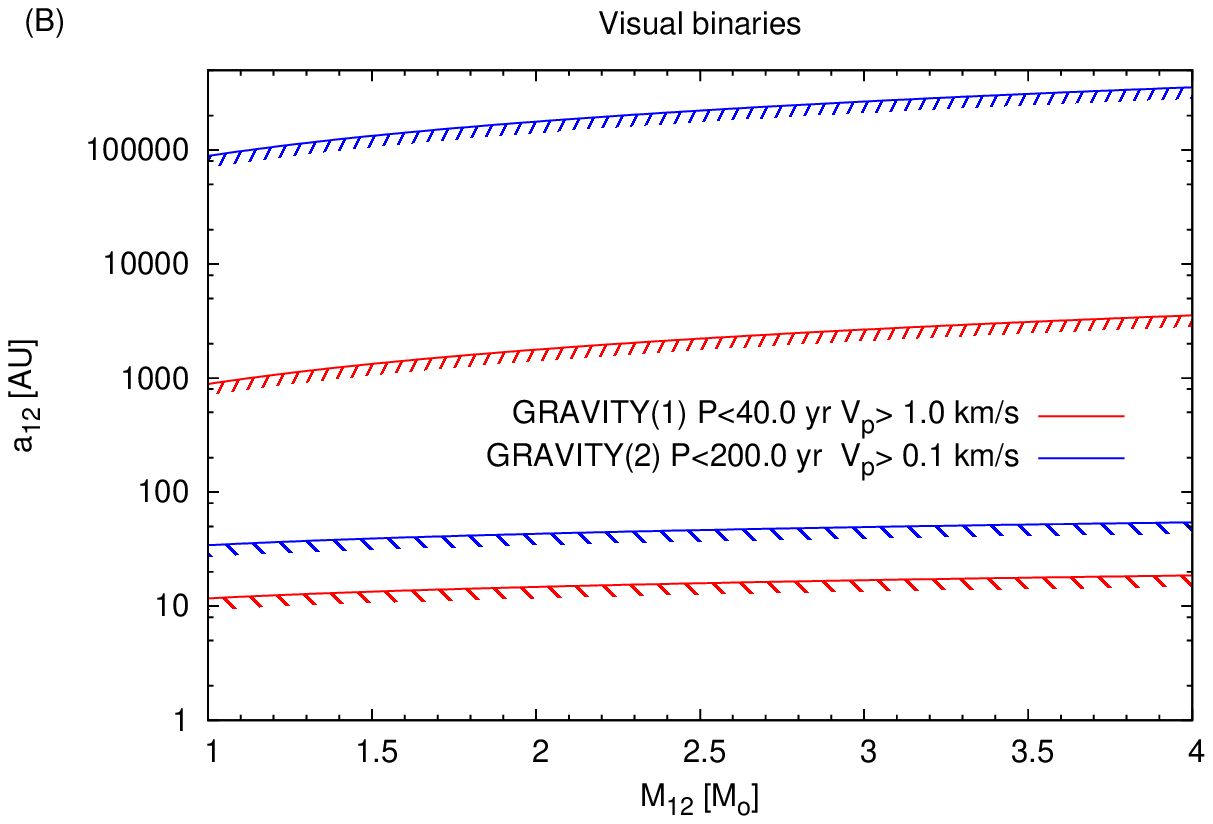}\tabularnewline
\end{tabular}\caption{\label{f:VP} (A) The detectability limit on the binary mass $M_{12}$
and sma $a_{12}$ for spectroscopic / astrometric binaries, for the
current period and velocity limits of SINFONI ($P<3$ yr and $V_{r}>20\,\mathrm{km\, s^{-1}}$
for $e=0$ and $\sin i=1$) and for future limits of GRAVITY (Case
1: conservative $P<4$ yr and $V_{p}>1\,\mathrm{km\, s^{-1}}$. Case
2: optimistic $P<20$ yr and $V_{p}>0.1\,\mathrm{km\, s^{-1}}$, for
$e=0$), in the low-mass range where binaries are long-lived enough
to probe relaxation (Eq. \ref{e:TMSdetect}). Binaries are detectable
in the region that is below the lower of the $\max\Pb$ upper bound
(coarse hashes) and the $\min V$ upper bound (fine hashes) for each
of the three cases. (B) The same as above, for visual binaries observed
by GRAVITY.}
 
\end{figure}

Observations of the GC with SINFONI have been on-going for $T_{\mathrm{obs}}\sim10$
years now, mostly focused on the central 0.2 pc, and spaced between
several days to several years. This limits currently detectable binaries
to periods of at most $\max\Pb=T_{\mathrm{obs}}/2\sim5\,$ yr, and
realistically probably to $\lesssim3$ yr. The joint constraints of
$\max\Pb$ and $\min V_{r}$ (Eqs. \ref{e:Pdetect}, \ref{e:vdetect})
translate to joint constraints on the binary mass and sma,

\begin{eqnarray}
a & < & a_{P}=\sqrt[3]{G\Mb(2\pi\max\Pb)^{2}}\,,\nonumber \\
a & < & a_{V_{r}}=\left.G\Mb\sin^{2}i\right/\min V_{r}^{2}\,,\label{e:aPVr}
\end{eqnarray}
where $a_{V_{r}}$ is estimated conservatively for a circular orbit.

The upcoming AO-assisted IR interferometer GRAVITY \citep{eis+11}
will allow the detection of binaries with separations up to several
AU. The instrument will provides two detection regimes: Detection
by fully resolving the two stars (``visual binaries''), and detection
by the shift of the binary's photo-center (``astrometric binaries'').
GRAVITY's astrometry will reach a precision of $10\,\mu\mathrm{as}$
(0.08 AU). Over a 4 year period this will allow the measurement of
proper velocities and accelerations at precisions of $\mu\sim10\,\mu\mathrm{as}\,\mathrm{yr^{-1}}$
($V_{p}\sim0.4\,\mathrm{km\, s}{}^{-1}$) and $\dot{\mu}\sim10\,\mu\mathrm{as\, yr^{-2}}$
($\mathrm{d}V_{p}/\mathrm{d}t\sim1.25\times10^{-8}\,\mathrm{km\, s^{-2}}$),
respectively, on sources as faint as $K_{S/N}\sim16$ \citep{bar+08,eis+11}.
Over 10 years these will improve to $\mu\sim2\,\mu\mathrm{as}\,\mathrm{yr^{-1}}$
($V_{p}\sim0.08\,\mathrm{km\, s}{}^{-1}$)\texttt{\textbf{ }}and $\dot{\mu}\sim2\,\mu\mathrm{as\, yr^{-2}}$
($\mathrm{d}V_{p}/\mathrm{d}t\sim2.5\times10^{-9}\,\mathrm{km\, s^{-2}}$).
We conservatively assume below $T_{\mathrm{obs}}=4\,\mathrm{yr}$,
but also consider the optimistic case of $T_{\mathrm{obs}}=20\,\mathrm{yr}$.

\paragraph{Visual binaries}

The high resolution of the VLTI allows to resolve binaries with separations
$>1\,\mathrm{mas}$ (8 AU at the distance of the GC). We will refer
to these binaries as visual binaries, since their separation can be
directly measured as in classical imaging. The typical orbital time
for a $\Mb=2\,\Mo$ binary with a separation of 8 AU is 16 yr; for
a separation of 25 AU it is 83 yr. This realistically implies coverage
of only a fraction of the orbit. However, this in itself will not
prevent the detection of such long-period binaries. Even in the dense
environment of the central cluster, the probability of a chance alignment
of two stars at a projected distance of $1\,\mathrm{mas}$ (8 AU)
from each other is ${\cal O}(10^{-6})$ and for $3\,\mathrm{mas}$
(25 AU) it is ${\cal O}(10^{-5})$ (Eq. \ref{e:NK}). Consequently,
pairs detected at separations of $\lesssim\mathrm{few\times10}$ AU
are true binaries at a very high confidence, and their detection can
already constrain the relaxation time even without orbital information.
Measurements of acceleration over arcs of the orbit can then be used
to derive the orbital parameters. Guided by such analysis of partial
orbits by \citet{cve+10}, we will assume here that orbits as long
as $\max\Pb=10T_{\mathrm{obs}}$ can be solved. The minimal detectable
plane-of-sky velocity amplitude $\min V_{p}$ then translates to the
additional constraint of 
\begin{equation}
a<a_{V_{p}}=\left.G\Mb\right/\min V_{p}^{2}\,,\label{e:aVp}
\end{equation}
where again $a_{V_{p}}$ is estimated conservatively for a circular
orbit.

Note that such long-period orbits somewhat relax the constraint on
the maximal mass useful for probing the dynamics %
\footnote{$\max\Ms\propto a_{12}^{2/7}$ since $\min T_{MS}\propto\Mb/\ab$
(Eq. \ref{e:TMSdetect}) and $T_{MS}(\Ms)\propto\Ms^{-5/2}$ for $\Ms\lesssim10\,\Mo$.%
}, since on such softer orbits even shorter-lived stars can still provide
interesting dynamical constraints on the dark cusp.

\paragraph{Astrometric binaries}

GRAVITY's novel astrometric mode will make it possible to detect unresolved
binaries by the astrometric shift of their common photo-center, provided
they have unequal fluxes. This requirement will most likely be satisfied
even for the fiducial near-equal mass red giant binaries that are
assumed in our analysis. Even a small difference in the mass leads
to a large difference in luminosity, and even more so, a small difference
in the post-main sequence evolutionary stage, as can be demonstrated,
for example, by the median visual magnitude difference of $\Delta m\simeq0.65$,
observed in binaries of \citep{har+01}. Although the amplitude of
the photo-center shift is $\sim\Delta m$ times smaller than the true
orbital motion of the brighter component, it does contain additional
constraints on the orbit beyond those provided by spectroscopy alone.
The detailed methodology for dealing with astrometric binaries will
be discussed elsewhere. Here we simplify the analysis by ignoring
the photo-center shift, and treat the binary detectability limits
for the GRAVITY astrometric mode in a similar way as those of spectroscopic
binaries. Following the criteria suggested by \citet{har+01}, we
restrict ourselves to binaries with $\max\Pb=T_{\mathrm{obs}}$.

Figure (\ref{f:VP}) shows the range of binary mass and sma where
spectroscopic binaries are now detectable with SINFONI, and where
astrometric and visual binaries will be detectable in the future with
GRAVITY (for both the conservative and optimistic sets of assumptions),
focusing on the low-mass range that is relevant for testing the dynamical
state of the GC. The limiting factor in all cases is the maximal binary
period required for detection and orbital parameter derivation. For
spectroscopic SINFONI binaries and for astrometric GRAVITY binaries
in the conservative case, the binary sma is limited to $\max\ab\sim3\:\mathrm{AU}$.
We therefore conclude that in this case the big improvement in astrometric
sensitivity by GRAVITY will not increase discovery space by much,
but will rather allow much better determination of the orbital parameters.
GRAVITY's advantage will be most significant in the optimistic case,
where the monitoring extends over much longer periods, or for visual
binaries. In that case discovery space will increase to $\max\ab\sim10\:\mathrm{AU}$
for both astrometric binaries and visual binaries in the conservative
case, and up to $\max\ab\sim50\:\mathrm{AU}$ for visual binaries
in the optimistic case. Thus, the combination of various spectroscopic
and interferometric techniques will allow the detection of binaries
in the GC with separations ranging from a few stellar radii to several
$10\,\mathrm{AU}$.

\subsubsection{The red giant binary population in the GC}

\begin{figure*}
\noindent \begin{centering}
\begin{tabular}{c}
\includegraphics[width=0.85\textwidth]{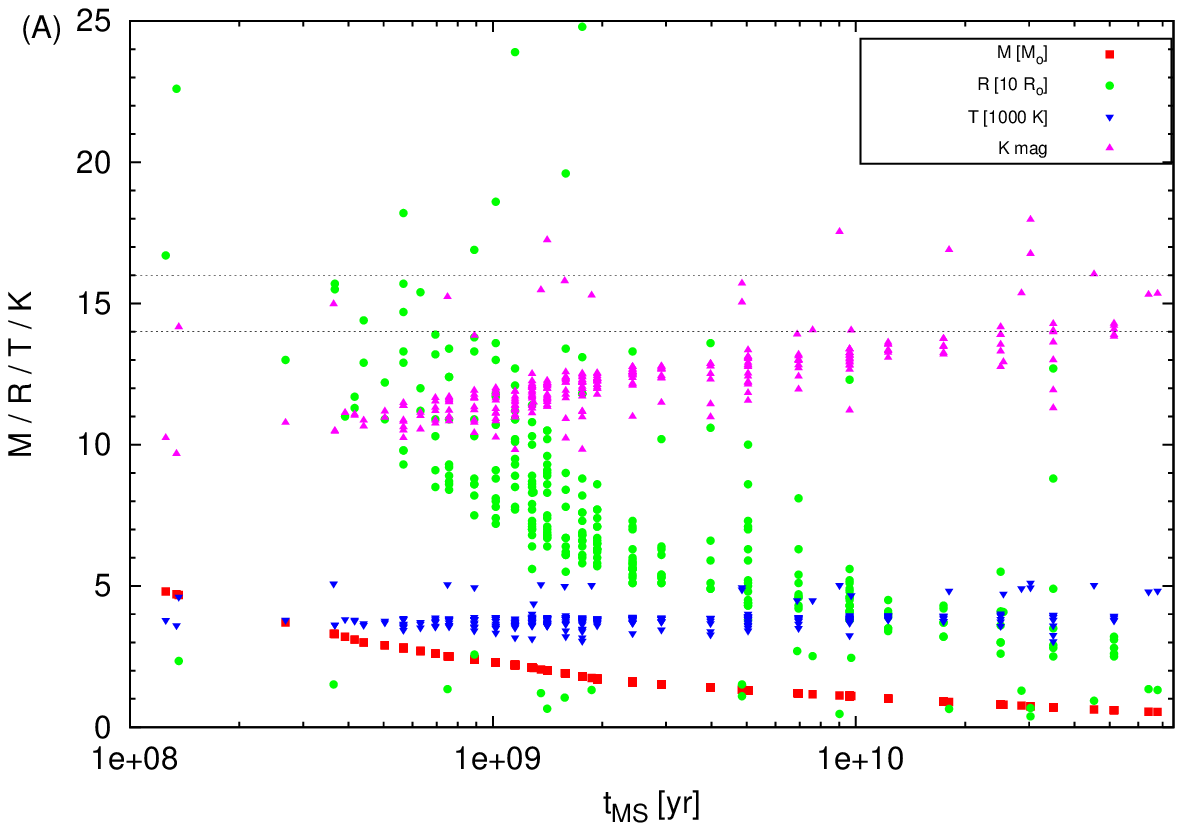}\tabularnewline
\includegraphics[width=0.85\textwidth]{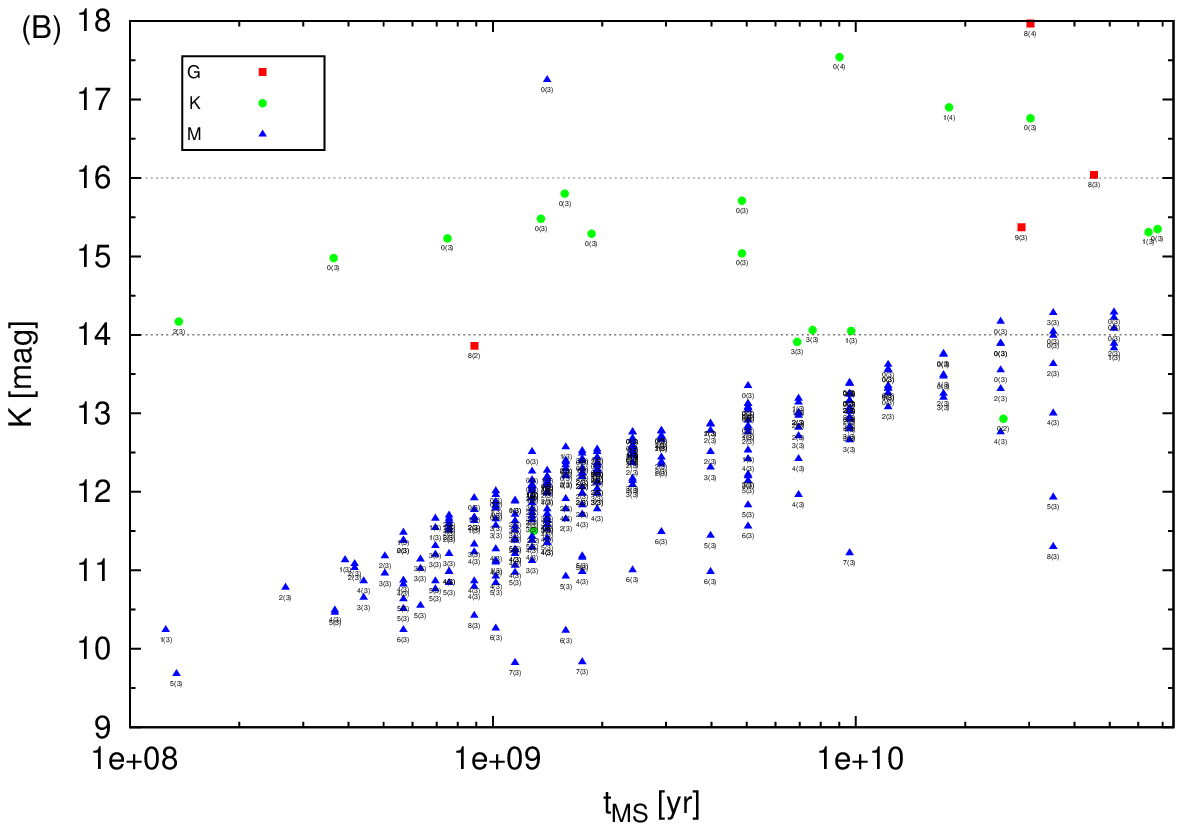}\tabularnewline
\end{tabular}
\par\end{centering}

\caption{\label{f:RG}The observed and deduced properties of samples of G,K
\citep{gra+82} and M \citep{dum+98} low-mass red giants, translated
to the GC ($\mathrm{DM}=14.6$, $A_{K}=2.7\,\mathrm{mag}$). The main
sequence lifetimes were adapted from the \citet{sch+92a} Solar metallicity
stellar tracks between $\Ms=0.8\,\Mo$ to $120\,\Mo$ and extrapolated
to lower masses by $T_{MS}(\Ms)=T_{MS}(0.8\,\Mo)(\Ms/0.8\,\Mo)^{-2.5}$.
(A) The red giant mass, radius, effective temperature and $K$ magnitude
in the GC. (B) The $K$ magnitudes at the GC for the different spectral
types (with the designation of the sub-class and luminosity class
in the format ``sc(lc)''). The $K_{\mathrm{S/N}}=14,16$ thresholds
are over-plotted for convenience. }
\end{figure*}

As argued above, the constraints set by current and forthcoming observational
capabilities reduce to three primary requirements: low-mass red giants
with $M_{12}<4\,\Mo$, a maximal magnitude $K\lesssim14$ for SINFONI
or $K\lesssim16$ for GRAVITY, and a maximal sma $\ab\lesssim3$ AU
for SINFONI or $a_{12}\lesssim\mathrm{few}\times10$ AU for GRAVITY.
We now estimate the number of detectable low-mass red giant binaries
in the GC (the ``discovery space'') from the observed $K$ luminosity
function in the GC, for the \emph{null hypothesis, }namely the assumption
that the effects of dynamical evolution on the binaries' properties
are negligible, and therefore binaries in the GC have the low-mass
binary fractions observed in dense Galactic clusters and the low-mass
binary period distribution observed in the field. 

The $K$ luminosity function of the red giants in the inner $\sim1/2$
pc is well approximated by \citep{ale+99a,bar+10} 
\begin{equation}
N(<K,\, R<1/2\,\mathrm{pc})\simeq10^{-2.4+0.35K}\,,\label{e:NK}
\end{equation}
 which allows the observed total surface number density of red giants
in the GC \citep{buc+09,bar+10} to be scaled to a given magnitude
limit. Assuming the observed and deduced range of binary fractions
in galactic clusters and that of massive binaries in the GC, $F_{2}\sim0.05-0.5$,
we estimate that there may be $N_{2}\sim15$--$150$ low-mass binaries
among the $N_{K}\sim300$ $K<14$ red giants that are observed within
1/2 pc of the MBH, and $N_{2}\sim80$--$800$ binaries among the $N_{K}\sim1600$
$K<16$ red giants there. 

A complete survey of low-mass $M_{12}\simeq2\,\Mo$ field binaries
indicates a log-normal orbital period distribution with $\left\langle \log_{10}\Pb\right\rangle =4.8$
and $\sigma_{\log_{10}\Pb}=2.3$ for $P_{12}$ in days (corresponding
to a median period of $\simeq180$ yr) \citet{duq+91}. Assuming that
the shorter period tail of this distribution applies also in the GC%
\footnote{A truncation of the long period tail of the distribution by evaporation
will increase the normalization constant relative to the untruncated
one. The normalized \citet{duq+91} distribution therefore provides
an under-estimate of the actual fraction of shorter-period binaries.%
}, then the fraction of binaries with $\ab<2.5$ AU ($\Pb<2.8$ yr)
is at least $F_{P}=0.22$, that of binaries with $\ab<8$ AU ($\Pb<16$
yr) is at least $F_{P}=0.33$, and that of binaries with $\ab<25$
AU ($\Pb<88$ yr) is at least $F_{P}=0.45$.. The number of potential
targets in each of these classes is $N_{2RG}=F_{P}F_{2}N_{K}$. \emph{We
therefore estimate that in the absence of strong dynamical evolution,
 there may exist in the central 1/2 pc of the GC $N_{2RG}\sim3-30$
low-mass $K<14$, $\Pb<3$ yr binaries, $N_{2RG}\sim25$--$250$ low-mass
$K<16$, $\Pb<8$ yr binaries, and $N_{2RG}\sim50$--$500$ low-mass
$K<16$, $\Pb<88$ yr binaries. }

Figure (\ref{f:RG}) shows the observed and derived properties of
galactic G, K and M red giants \citep{gra+82,dum+98}, translated
to the GC distance and extinction, as function of $T_{MS}$, where
the main-sequence lifespan was matched by mass to the $0.8-120\,\Mo$
Solar metallicity stellar tracks of \citet{sch+92a} and extrapolated
below $0.8\,\Mo$ assuming the low-mass limit relation $T_{MS}\propto\Ms^{-2.5}$
\citet[e.g.][]{han+04}. The $K$-magnitude was estimated from the
effective temperature. The plot shows that the effective temperature,
and hence spectral type, are nearly independent of the mass, and hence
$T_{MS}$, while the increase of the $K$-magnitude with the decrease
of mass is substantially smeared by large scatter. It is therefore
not possible to determine the stellar mass to better than a factor
of $\sim2$, and $T_{MS}$ to better than a factor of $\sim2^{2.5}\sim6$.
Table (\ref{t:RGs}) lists four low-mass red giants types that are
observed in the GC and serve below in examples of the dynamical constraints
that can be derived from low-mass red giant binaries. For demonstration
purposes, the mass range $0.8-2.0\,\Mo$ has been assigned in equal
increasing increments from the latest to the earliest types in the
table, and likewise the corresponding main-sequence lifespans in equal
decreasing logarithmic increments the range $2-12$ Gyr. However it
should be emphasized that for all these spectral types the uncertainty
in mass and stellar lifespan is such that $0.8\lesssim\Ms\lesssim2\,\Mo$
and $2\lesssim T_{MS}\lesssim12$ Gyr.

\begin{table}[t]
\caption{\label{t:RGs}Properties of typical red giants in the GC}

\noindent \centering{}%
\begin{tabular}{lccccc}
\hline 
\noalign{\vskip\doublerulesep}
Type  & $\Ms$ $^{1,3}$ & $\Rs$~$^{1}$ & $T_{\mathrm{eff}}$ & $K$~$^{2}$ & $T_{MS}$~$^{4}$\tabularnewline[\doublerulesep]
\hline 
\hline 
\noalign{\vskip\doublerulesep}
K2(III) & 2.0 $\Mo$ & 19 $\Ro$ & 4351 K & 15.1 & 2 Gyr\tabularnewline
K5(III) & 1.6 $\Mo$ & 33$\Ro$ & 4023K & 13.5 & 4 Gyr\tabularnewline
M0(III) & 1.2 $\Mo$ & 40$\Ro$ & 3914 K & 13.1 & 7 Gyr\tabularnewline
M2(III) & 0.8 $\Mo$ & 61$\Ro$ & 3695 K & 12.4 & 12 Gyr\tabularnewline[\doublerulesep]
\hline 
\noalign{\vskip\doublerulesep}
\multicolumn{6}{l}{{\footnotesize{$^{1}$~\citet{bel+99}}}}\tabularnewline
\noalign{\vskip\doublerulesep}
\multicolumn{6}{l}{{\footnotesize{$^{2}$~\citet{cox00}}}}\tabularnewline
\noalign{\vskip\doublerulesep}
\multicolumn{6}{l}{{\footnotesize{$^{3}$~Uncertain mass in the range $0.8-2\,\Mo$.}}}\tabularnewline
\noalign{\vskip\doublerulesep}
\multicolumn{6}{l}{{\footnotesize{$^{4}$~Uncertain main-sequence lifespan in the range
$2-12$ Gyr.}}}\tabularnewline
\hline 
\end{tabular}
\end{table}

\subsubsection{Dynamical constraints by GC red giant binaries}

\begin{figure}
\noindent \begin{centering}
\begin{tabular}{c}
\includegraphics[width=1\columnwidth]{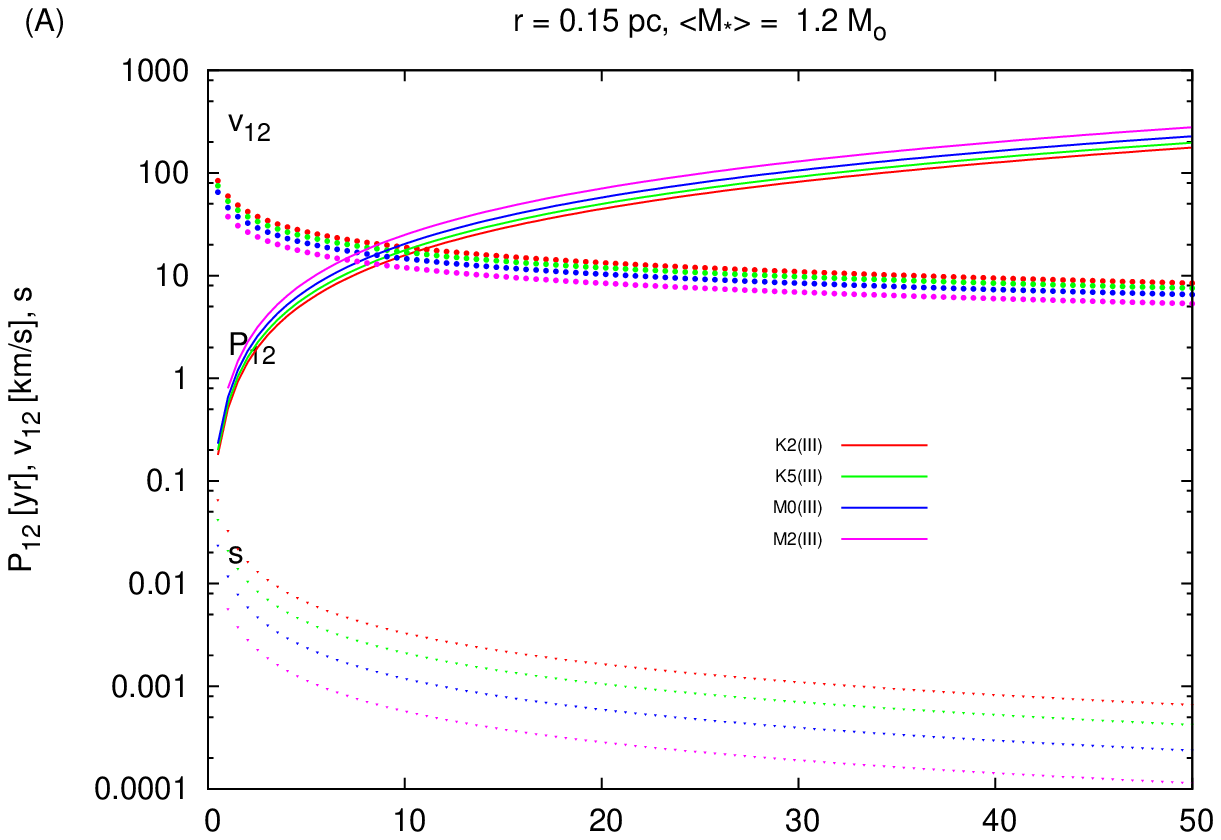}\tabularnewline
\includegraphics[width=1\columnwidth]{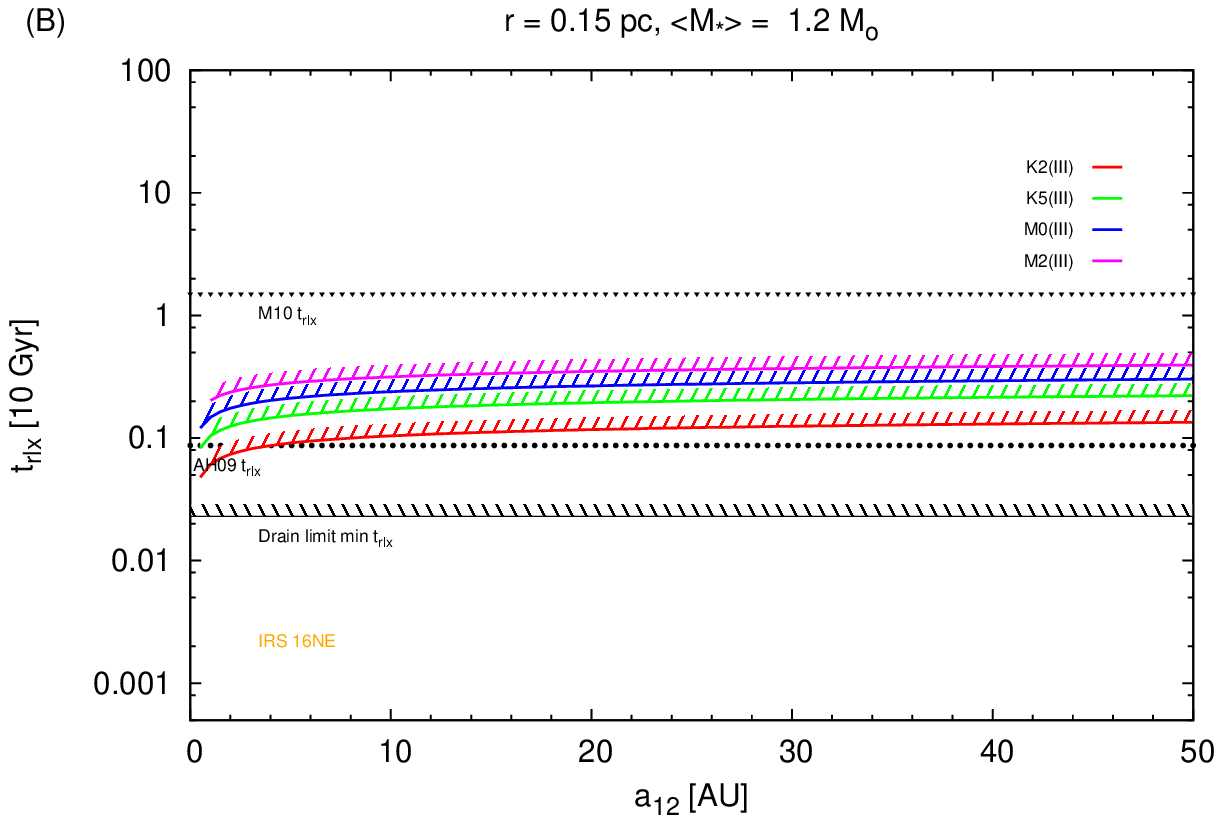}\tabularnewline
\end{tabular}
\par\end{centering}

\caption{\label{f:min_trlx_a}(A): The orbital period $\Pb$ reflex velocity,
$\vb$ and softness parameter $s$ (for $\alpha=7/4$ and $\eM=1.2\,\Mo$)
at $r=0.15$ pc of 4 types of equal mass binaries (Table \ref{t:RGs}),
as function of the binary sma in the range $\ab>2\Rs$. Over most
of the range of interest $\Pb$ is short enough and $\vb$ typically
high enough for detection. (B): The lower bound on the relaxation
timescale, $\mtr$, which could be inferred from the discovery of
a binary at $r=0.15$ (Eq. \ref{e:mintr}). $\eM=1.2\,\Mo$ is assumed
based on the GC model of \citet{ale+09}. The allowed values of $\tr$
extend above the hashed lines. Three theoretical predictions for $\tr(0.15\,\mathrm{pc)}$
are shown: The Drain Limit \citep{ale+04}, which provides a lower
bound on viable steady-state models, the mass-segregated GC model
of \citet{ale+09}, which is similar to that of \citet{fre+06}, and
the post-merger ``scoured'' GC model of \citet{mer10}. Also shown
is the bound provided by the recently discovered IRS 16NE. A model
that predicts a relaxation time that lies substantially \emph{below}
the lower limit provided by the binaries is ruled out (see discussion
in Sec. \ref{s:discussion}). }
\end{figure}

Figure (\ref{f:min_trlx_a} A) shows the binary period, the typical
reflex velocity (the actual velocity will depend on projection and
the eccentricity), and the softness parameter (assuming $\alpha=7/4$
and $\eM=1.2\,\Mo$, cf Figure \ref{f:meanM}), as function of binary
sma above the contact limit $\ab>2\Rs$, up to $\ab=50$ AU. 

\begin{figure}
\noindent \centering{}%
\begin{tabular}{c}
\includegraphics[width=1\columnwidth]{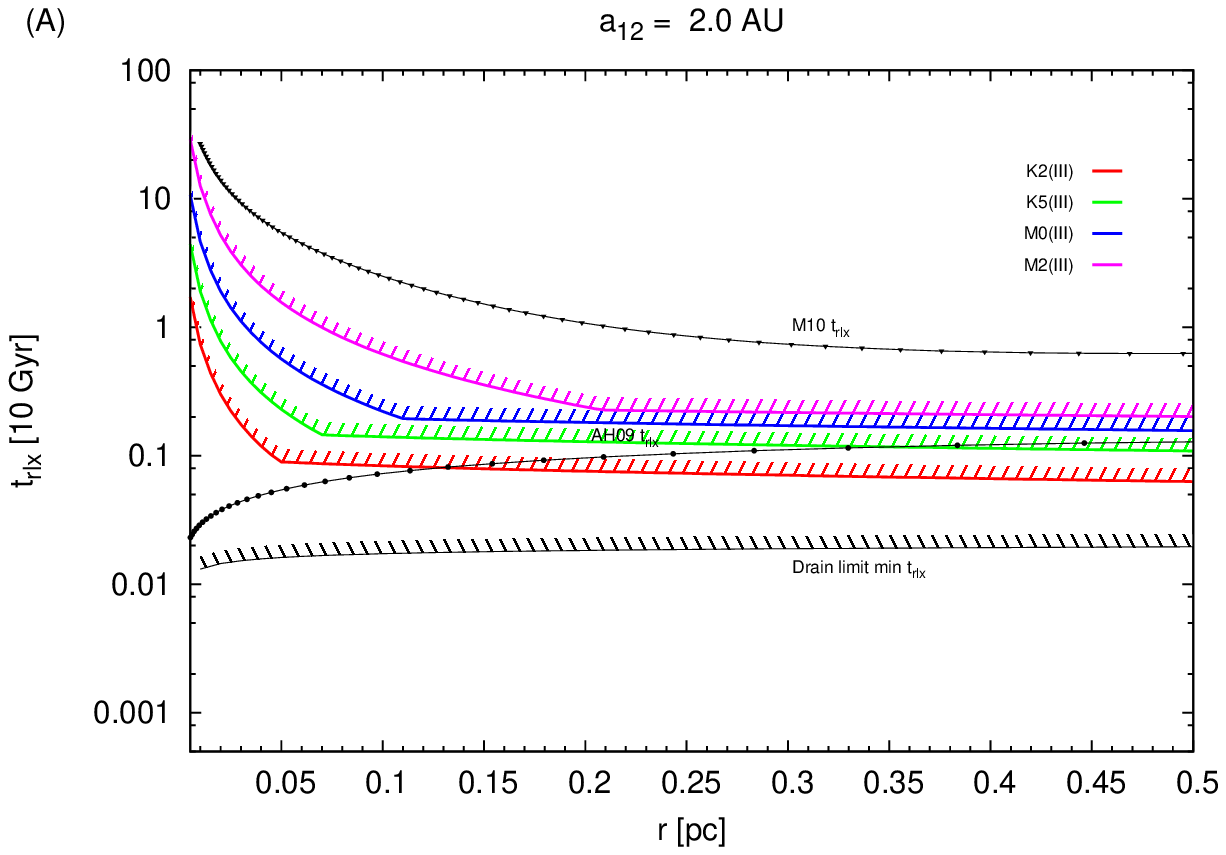}\tabularnewline
\includegraphics[width=1\columnwidth]{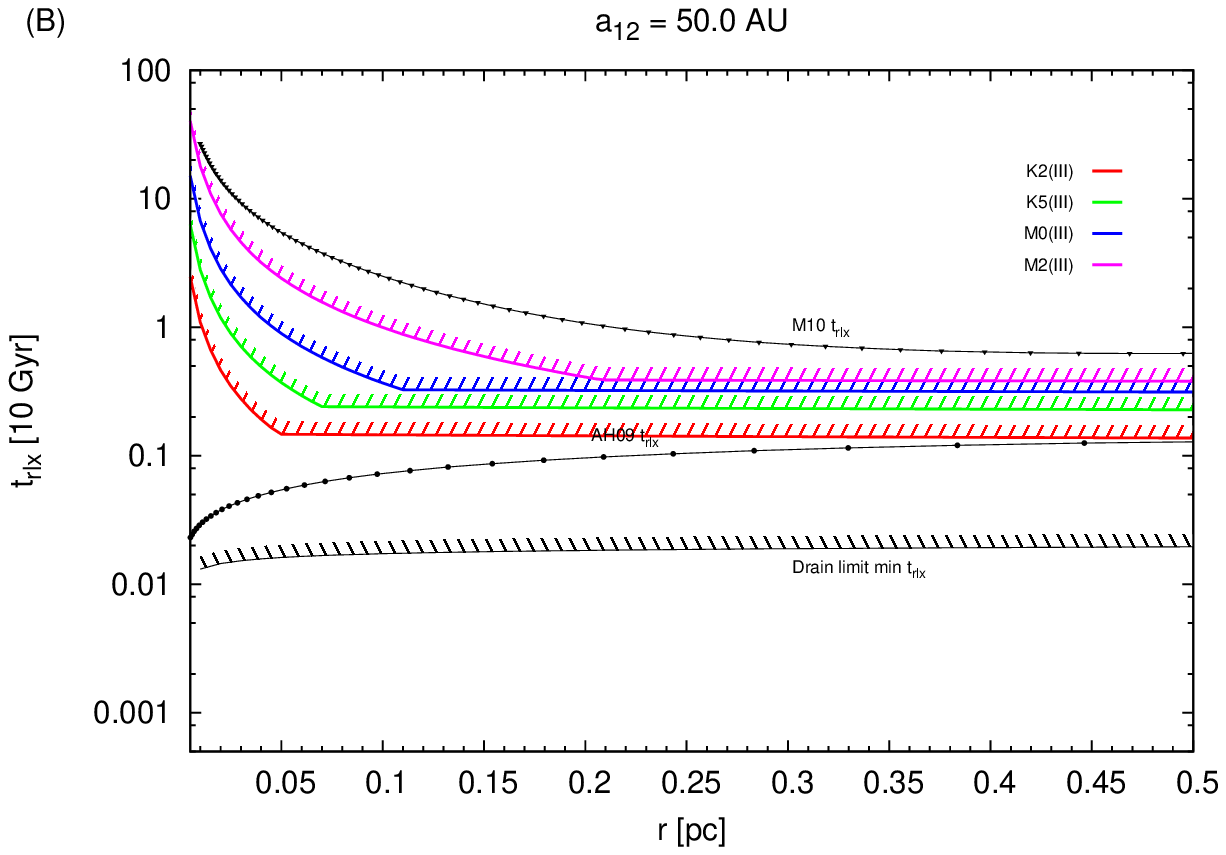}\tabularnewline
\end{tabular}\caption{\label{f:min_trlx_r}Same as in Figure (\ref{f:min_trlx_a} B), but
for a fixed binary sma, as function of distance $r$ from the MBH,
and assuming $\eM=0.88(r/1\,\mathrm{pc)^{-0.12}}$. based on the GC
model of \citet{ale+09} (Figure \ref{f:meanM}). (A) A spectroscopic
/ astrometric binary with $\ab=2$ AU. (B) A visual binary with $\ab=50$
AU.}
\end{figure}

Figure (\ref{f:min_trlx_a} B) compares the lower limits on the relaxation
time that can be derived by the detection of such binaries as function
of the sma, estimated for $r=0.15$ pc, to the various theoretical
models. The results show that the detection of such binaries there
can rule out that the system is at the drain limit, and can significantly
constrain mass-segregated models of the GC. Similarly, Figure (\ref{f:min_trlx_r})
shows the lower limits of these binaries at $\ab=2$ and $50$ AU,
as function of the distance from the MBH. The results show that because
the relaxation times in relaxed models typically decrease toward the
center, finding binaries there will provide stronger constraints on
the models. 

\section{Discussion and summary}

\label{s:discussion}

The detection of long-period binaries in the GC can provide lower
limits on the local 2-body relaxation timescale, and correspondingly,
upper limits on the number density of the dark cusp that is predicted
to exist there. However, since there is no compelling theoretical
prediction for the formation or capture rate of binaries there, the
converse is not true: the absence of binaries \emph{can not }prove
the existence of a high density cusp---it is merely consistent with
it.

We have shown that low-mass long-period binaries are already detectable
among the brighter red giants in the GC. We adopt the observed statistics
of such binaries in dense clusters and their period distribution in
the field as the \emph{null hypothesis} (binary formation in the GC
is not special, and there is no evaporation), and use these to estimate
the number of observable low-mass, long period binaries. Assuming
the null hypothesis, and taking into detailed account the VLT's existing
and forthcoming instrumentation and the observational and data analysis
constraints, we estimate that among the brighter $(K<14$) red giants
in the central 1/2 pc of the GC there could be $N_{2RG}\sim3-30$
spectroscopic binaries currently detectable by the integral field
spectrometer SINFONI/VLT and $N_{2RG}\sim50-500$ fainter ($K<16$)
astrometric and visual binaries that could be detected by the forthcoming
AO-assisted IR interferometer GRAVITY/VLT. The discovery potential
will increase as observations become deeper, the observing cadence
higher, and the monitoring baseline longer. This requires a specifically
planned long-term campaign. However, even the presently available
data may already include a few long-period binaries. 

It is quite likely that the null hypothesis is unrealistic. There
are reasons to believe that star formation and binary formation in
the unique environment near a MBH proceed quite differently than in
less extreme environments. The initial fraction of binaries in the
GC could be either much higher or much lower. For example, the dynamics
of binary formation in a fragmenting accretion disk possibly leads
to a very rapid hardening and merging of binaries \citep{bar+11}.
Conversely, the expected high density of compact remnants near the
MBH, presumably responsible to the over-abundance of low-mass X-ray
binaries there \citep{mun+05} suggests the intriguing possibility
of exotic binaries composed of a red giant and a stellar mass black
hole, which were not considered here. 

Since evaporation is a statistical process, robust limits on the dynamical
state of the GC require a sample of low-mass long-period binaries.
Our estimates of the number of detectable binaries indicate that a
dedicated search can plausibly yield at least a small sample, if some
binaries do indeed survive in the GC.

Binaries can also be destroyed by processes other than evaporation.
This will not affect our method for constraining the dark cusp, which
draws conclusions only from those binaries that do survive. This can
be stated formally as follows: assume that that are several competing
binary destruction processes, which singly would limit a binary's
lifetime to $t_{i}(\{p_{i}\})$, where $i=0,1,2,\ldots$ labels the
process (evaporation is process 0), and $\{p_{i}\}$ is the set of
physical parameters that process $i$ depends on (e.g. $\n$, $\eM$,
$\eMM,\ldots$), which can be constrained by the detection of surviving
binaries. The existence of a binary with age $\Tb$ then sets the
limit $\Tb<\min_{i}t_{i}\le t_{0}=t_{\mathrm{evap}}$ irrespective
of the relative efficiencies of the processes. However, the presence
of multiple binary destruction channels is relevant in that it decreases
the chances of finding any usable binaries. We now list some of the
competing destruction channels, without committing to specific rates
or rank order. We have argued \citep{pfu+13} that the tidal break-up
of a binary by the MBH occurs on a timescale of a $\mathrm{few}\times\tr\gg\tau_{\mathrm{evap}}$
(Eq. \ref{e:trlx-tevap}). Tidal breakup driven by incoherent 2-body
relaxation will therefore not compete with evaporation, except possibly
very near the MBH \citep{hop09}. On the other hand, breakup driven
by resonant relaxation could be more efficient, especially very near
the MBH \citep{rau+96,hop+06a}, where it could destroy binaries at
a rate faster than 2-body relaxation replenishes them \citep[e.g.][]{mad+11}.
In addition, accelerated binary mergers can be induced by Kozai perturbations
by the MBH. This effect is expected to be relevant for the subset
of high-inclination orbits close to the MBH \citep{ant+10,ant+12}.\texttt{\textbf{ }}

We have shown how the properties of a detected binary: location, mass,
period, main-sequence lifespan, and the observed and deduced properties
of the GC, the MBH mass and the stellar velocity dispersion, are translated
to a lower bound on the 2-body relaxation time $\tr$ (\ref{e:mintr})
and an upper bound on the second moment of the mass distribution,
$\n\eMM$ (\ref{e:maxnM2}), \emph{independently of the uncertainties
in the binary fraction in the GC}. In concise form, these bounds can
be expressed in terms of the observed or derived binary properties:
The main-sequence lifetime $T_{12}$, the orbital velocity scale $\vb^{2}=G\Mb/\ab$,
the Coulomb logarithm $\log\Lambda_{12}$ (Eq. \ref{e:L12}), and
the maximal evolution ratio $S_{h}$ (Eq. \ref{e:Sh}), through the
combination 
\begin{equation}
X_{12}(r)=\frac{T_{12}\log\Lambda_{12}}{\vb^{2}S_{h}(r)}\,.
\end{equation}
Together with assumed model values for $\alpha$, the logarithmic
slope of the stellar cusp, $\eM$, the mean stellar mass, and the
observed or modeled velocity dispersion $\sigma_{\star}$ (Eq. \ref{e:sig2}),
the bounds on the relaxation time and stellar density are 
\begin{eqnarray}
t_{\mathrm{rlx}}(r) & > & 1.4\frac{\s^{2}(r)}{\log\left(\Mbh/\eM\right)}X_{12}(r)\,,\nonumber \\
\n\eMM & < & 0.24\frac{\s(r)}{G^{2}}X_{12}^{-1}(r)\,.
\end{eqnarray}
We have shown that the detection of such binaries could rule out the
possibility of a maximally dense relaxed stellar cusp at the drain
limit, and constrain more realistic relaxed dynamical models of the
GC. 

The key issue, and the main source of uncertainty, is to determine
how long the binary spent in or near its observed location. The actual
age (as opposed to the lifespan) of red giants can be estimated from
the main-sequence lifespan of their progenitors, $T_{MS},$ which
in principle could provide $\Tb\sim T_{MS}$ to within 10\% (the typical
relative duration of the post-main sequence phase). This is different
from the case of young massive stars, whose age can only be bound
by their main-sequence lifespan, $\Tb<T_{MS}(\Ms)$, unless they can
be associated with the stellar disk, whose age is known, $\Tb\simeq(6\pm2)\,\mathrm{Myr}$. 

We have argued that local tidal or 3-body formation in the GC is negligible
(Sec. \ref{ss:3body}), and therefore the binaries orbiting the Galactic
MBH are either primordial and local, or primordial and captured by
a $4$-body tidal interaction between a hierarchical triple and the
MBH \citep{per09b}, in a variation of the \citet{hil88} mechanism.\texttt{\textbf{ }}The
fraction of such captured binaries among the local binaries is uncertain.
They will spend a substantial fraction of the post-capture time on
very eccentric orbits, until these are randomized \citep{per+09b},
which may provide a way to identify them if they are relatively recently
captured. In any case, under the plausible assumption that both their
time of formation and their time of capture are random and distributed
uniformly in time over their main-sequence lifespan $T_{MS}$, then
their mean time in the GC is $\left\langle \Tb\right\rangle \sim T_{MS}/4$,
which is within the expected uncertainty in the determination of the
main-sequence lifetimes of the low-mass red giant binaries (table
\ref{t:RGs}). 

The analysis presented here makes several simplifying assumptions,
which should be relaxed and calculated in more detail. (1) The estimates
here are based on the present-day location of the binary in the GC,
without orbital averaging. For long-lived binaries, the evolution
of the binary's orbit around the MBH is not negligible and should
be also taken into account, as should the evolution of the nuclear
cluster itself. (2) The statistical fluctuations in the stochastic
process of evaporation are not taken into account. Evaporation should
be expressed by survival probabilities, rather than by a single timescale.
(3) The estimates are based on the present-day binary parameters $\ab$
and $\Mb$ and take into account only the dynamical evolution of soft
binaries due to external interactions with stars. However, these parameters
also evolve internally because of post-main sequence mass-loss, which
is neglected here. Issues (1) and (2) in particular should be addressed
by $N$-body modeling of the evaporation process.

Finally, it should be stressed that binary evaporation is insensitive
to relaxation by spatially extended massive perturbers (e.g. giant
molecular clouds, or dense clusters), which can shorten the relaxation
time by many orders of magnitude \citep{per+07}, since these will
not widen the binary, but rather scatter its center of mass. Thus,
$\max\n\eMM$ is a limit on the point-like masses in the system (stars,
compact remnants and possibly also intermediate-mass BHs), and $\min\tr$
is a lower limit on the \emph{stellar} 2-body relaxation timescale
alone. The actual relaxation time due to both stars and extended massive
perturbers may be shorter still.

\noindent \begin{center}

\par\end{center}

\acknowledgements{T.A. acknowledges support by ERC Starting Grant No. 202996, DIP-BMBF
Grant No. 71-0460-0101, and the I-CORE Program of the PBC and ISF
(Center No. 1829/12). T.A. is grateful for the hospitality of the
Kavli Institute for Theoretical Physics, UC Santa Barbara, where this
work was completed. This research was supported in part by the National
Science Foundation under Grant No. NSF PHY11-2591.}

\bibliographystyle{apj}

\end{document}